\documentclass[%
 reprint,
 amsmath,amssymb,
 pre,
 aps,
 superscriptaddress,
 twocolumn
]{revtex4}

\usepackage{graphicx}% Include figure files
\usepackage{dcolumn}% Align table columns on decimal point
\usepackage{bm}% bold math
\usepackage{amsmath,amssymb,mathtools}

\usepackage{color}
\definecolor{darkblue}{rgb}{0,0,0.6}
\definecolor{darkred}{rgb}{0.6,0,0}
\definecolor{darkgreen}{rgb}{0,0.6,0}
\usepackage[colorlinks=true,urlcolor=darkblue,citecolor=darkblue,linkcolor=darkred,hyperfootnotes=false]{hyperref}

\begin{document}

%\preprint{APS/123-QED}

\title{Effects of the self-propulsion parity on the efficiency\\of a fuel-consuming active heat engine}% Force line breaks with \\

\author{Yongjae Oh}
\author{Yongjoo Baek}%
 \email{y.baek@snu.ac.kr}
\affiliation{%
Department of Physics and Astronomy \& Center for Theoretical Physics, Seoul National University, Seoul 08826, Republic of Korea
 %\\This line break forced with \textrmbackslash\textrmbackslash
}%
\date{\today}

%\affiliation{Second institution and/or address
%\\This line break forced with \\}%
%\affiliation{
 %Third institution, the second for Charlie Author
%\affiliation{%
 %This line break forced with \textrmbackslash\textrmbackslash
%}%



\begin{abstract}
We propose a thermodynamically consistent, analytically tractable model of steady-state active heat engines driven by both temperature difference and a constant chemical driving. While the engine follows the dynamics of the Active Ornstein-Uhlenbeck Particle, its self-propulsion stems from the mechanochemical coupling with the fuel consumption dynamics, allowing for both even- and odd-parity self-propulsion forces. Using the standard methods of stochastic thermodynamics, we show that the entropy production of the engine satisfies the conventional Clausius relation, based on which we define the efficiency of the model that is bounded from above by the second law of thermodynamics. Using this framework, we obtain exact expressions for the efficiency at maximum power. The results show that the engine performance has a nonmonotonic dependence on the magnitude of the chemical driving, and that the even-parity (odd-parity) engines perform better when the size of the engine is smaller (larger) than the persistence length of the active particle. We also discuss the existence of a tighter upper bound on the efficiency of the odd-parity engines stemming from the detailed structure of the entropy production. 
\end{abstract}

\maketitle

%\tableofcontents

\section{Introduction}

Formulation of the macroscopic irreversibility in terms of the entropy production and its application to the upper bound on the efficiency of heat engines was a cornerstone in the development of thermodynamics. The Clausius relation, which relates energy exchanges with thermal reservoirs to the change of entropy, provides a systematic way to describe the fundamental limitations of how efficient an engine can be, namely the Carnot efficiency, which is the maximum efficiency attainable by a quasistatic process. But the original Clausius relation is applicable only to quasistatic processes, during which systems stay close to equilibrium.

Since various natural and artificial engines operate far from equilibrium to achieve finite power, modern thermodynamics has focused on developing a systematic framework for describing the irreversibility of such systems. More recently, with the development of technologies for observing and controlling nanoscale systems, microscopic engines subject to nonnegligible thermal and athermal fluctuations have been constructed. The development of stochastic thermodynamics~\cite{seifert2012stochastic} proved to be a crucial step towards describing such system. The theory provides a systematic method for deriving a host of inequalities describing the irreversibility of a broad range of systems, including the systems driven by finite-time protocols with nonnegligible microscopic fluctuations.

An important challenge of this field is to describe the performance of engines composed of a single active particle. An active particle maintains its direction of motion by converting its stored energy into a propulsion force determined by its internal degrees of freedom~\cite{marchetti2013hydrodynamics, bechinger2016active,ramaswamy2017active}. Examples can be found in various natural and artificial systems, such as flocks of birds, school of fish, swimming bacteria, Janus particles, and colloidal rollers. They have been extensively studied for their novel far-from-equilibrium collective phenomena such as flocking, phase separation~\cite{cates2015motility,tjhung2018cluster,fausti2021capillary}, current rectification~\cite{dileonardo2010bacterial}, and formation of topological defects~\cite{yeomans2014active,saw2017topological} But recent studies have also investigated engines composed of such active particles, namely {\em active heat engines}~\cite{fodor2021active,pietzonka2019autonomous,ekeh2020thermodynamic,krishnamurthy2016micrometre, lee2020brownian,martin2018extracting, holubec2020active,holubec2020underdamped,kumari2020stochastic,speck2022efficiency}.

Active heat engines are distinct from the ordinary heat engines ({\em passive heat engines}) in that they do not require temperature difference to operate. Positive work can be extracted from such engines even in isothermal environments by constant protocols imposing a nonequilibrium steady state~\cite{pietzonka2019autonomous} or by cyclic protocols involving other control parameters~\cite{martin2018extracting}. Even periodic manipulation of the potential alone is enough for persistent work extraction~\cite{ekeh2020thermodynamic}. This makes active heat engines an ideal candidate for designing nanomachines or micromachines operating in living systems, whose temperature does not vary much.

It is natural to ask how to define efficiency of such engines. According to the standard methods of stochastic thermodynamics, for isothermal active heat engines, the {\em active work}, {\em i.e.}, the work done by the self-propulsion force, yields an upper bound on the extractable work. Thus, the ratio between the extracted work and the active work, bounded from above by $1$, was used as the definition of efficiency in \cite{ekeh2020thermodynamic, pietzonka2019autonomous,speck2022efficiency}. In case the fuel consumption is tightly coupled to the motion of the active particle, the active work is equivalent to the {\em chemical work}, as was the case in \cite{pietzonka2019autonomous}. We also note that the standard definition of efficiency for {\em molecular motors} in the literature~\cite{schmiedl2008efficiency,parmeggiani1999energy,pietzonka2016universal} is the ratio between the extracted work and the chemical work, whose upper bound is also $1$.

Meanwhile, active heat engines operating between different temperatures have also been extensively studied~\cite{krishnamurthy2016micrometre,holubec2020active,holubec2020underdamped,kumari2020stochastic,datta2022second}. Interest in such engines was sparked by the experiment of an {\em active Stirling engine}, which used swimming bacteria confined in a laser trap to extract work via cyclic protocols~\cite{krishnamurthy2016micrometre}. The study reported that the ``efficiency'' of the engine, defined as the ratio between the extracted work $W_\mathrm{out}$ and the ``heat'' absorbed by the system from the hot reservoir, defined as $\Delta E - W_\mathrm{out}$ for the change of internal energy $\Delta E$ during the process, can surpass the Carnot efficiency (``super-Carnot behavior''). This behavior was attributed to the non-Gaussian statistics of the swimming bacteria~\cite{krishnamurthy2016micrometre} even in the harmonic optical potential. A theoretical work by \cite{lee2020brownian}, which revisited the experiment from the perspective of a steady-state engine simultaneously coupled to two reservoirs, attributed the super-Carnot behavior to the finite correlation time incurred by the swimming bacteria. The same definition of efficiency was also used in \cite{holubec2020active,holubec2020underdamped,kumari2020stochastic} for active heat engines operating between different temperatures.

While the definition of efficiency used in those studies is a straightforward generalization of the definition used for the efficiency of the Carnot engine, they lack any upper bound stemming from the second law of thermodynamics. This is because they do not distinguish between different components of the ``heat'' $\Delta E - W_\mathrm{out}$, which is actually a mixture of the proper heat from the reservoir and the chemical work done on the particle. These distinct types of energy flows may contribute differently to the irreversibility of the system, which should be clarified by explicitly modeling the dynamics of chemical degrees of freedom and applying the methods of stochastic thermodynamics.

Also related is the issue of quantifying how far from equilibrium active particles are. The dynamics of active particles are typically modeled at a phenomenological level, introducing a nonequilibrium driving that breaks the fluctuation-dissipation theorem at the particle level. A notable example is the Active Ornstein-Uhlenbeck Particle (AOUP)~\cite{szamel2014self,koumakis2014directed}, which is kept out of equilibrium by making the time scale of friction (assumed to be instantaneous) different from the correlation time of the athermal noise (assumed to be finite). Lacking a full energetic picture of how such nonequilibrium driving arises, the irreversibility of such apparent dynamics~\cite{fodor2016far} is disconnected from the energy flows~\cite{crosato2019irreversibility}. While the notion of apparent irreversibility is useful for characterizing whether an effective equilibrium description is possible for the {\em dynamics} of active particles, it does not describe the {\em thermodynamics} of how much energy is dissipated to maintain certain structures formed by active particles. Notably, the lack of a clear energetic picture of how self-propulsion arises has led to some controversy about the irreversibility of active particles, especially regarding whether the self-propulsion force should change sign under time reversal~\cite{mandal2017entropy,caprini2018comment,mandal2018mandal}. Now, the consensus is that both positive and negative signs (called {\em even} and {\em odd} parities, respectively) are equally possible, with the suitable parity to be determined by the detailed picture of how the self-propulsion arises~\cite{dabelow2019irreversibility, crosato2019irreversibility, shankar2018hidden, fodor2022irreversibility}.

\begin{figure}[b]
\includegraphics{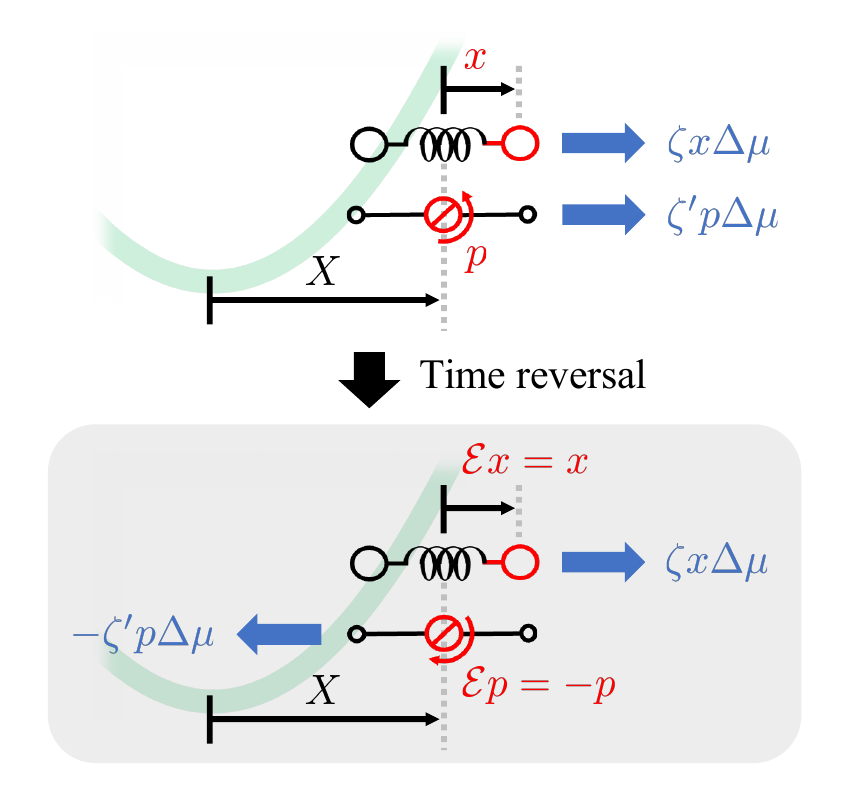}% Here is how to import EPS art
\caption{\label{fig:schematic}
An illustration of two different types of AOUPs with even-parity and odd-parity self-propulsion forces. Here, 
$X$ denotes the position of the AOUP, and $x$ ($p$) is an internal degree of freedom which keeps (changes) its sign under time reversal. The dimers propel themselves (as indicated by thick horizontal arrows) thanks to the constant chemical driving $\Delta\mu$. The behaviors of the self-propulsion forces under time reversal are shown inside the grey box, where $\mathcal{E}$ is the time-reversal operator. More detailed descriptions of the model are provided in Sec.~\ref{sec:chem_aoup}.}
\end{figure}

For these reasons, many studies have focused on constructing {\em thermodynamically consistent} descriptions of active particles~\cite{ramaswamy2017active,pietzonka2017entropy,speck2018active,gaspard2018fluctuating,dadhichi2018origins}, which aim for models that are detailed enough to relate the irreversibility of the stochastic dynamics to the energy flows, as done by the Clausius relation in the conventional thermodynamics. Those models introduce a constant chemical driving as the origin of self-propulsion and provide a coherent picture of how such mechanochemical coupling affects the dynamics of both mechanical and chemical degrees of freedom. However, application of this approach to a concrete description of the performance of an active heat engine is still lacking.

In this study, employing the {\em active dimer} framework used in \cite{dadhichi2018origins}, we construct a thermodynamically consistent, analytically tractable model of a fuel-consuming active heat engine, whose dynamics follows the AOUP with the self-propulsion stemming from a constant chemical driving. Both even-parity and odd-parity self-propulsion are considered, with minimal descriptions of the mechanochemical coupling for each situation, which allows us to properly distinguish between the heat and the chemical work. Then, applying the standard methods of stochastic thermodynamics, we derive the Clausius relations between the entropy production of the engine and the heat flows, which yields a definition of the engine efficiency properly bounded from above by the second law of thermodynamics. This allows us to systematically assess the performance of the active heat engine for self-propulsion force of both parities.

\begin{figure*}%[b]
\includegraphics{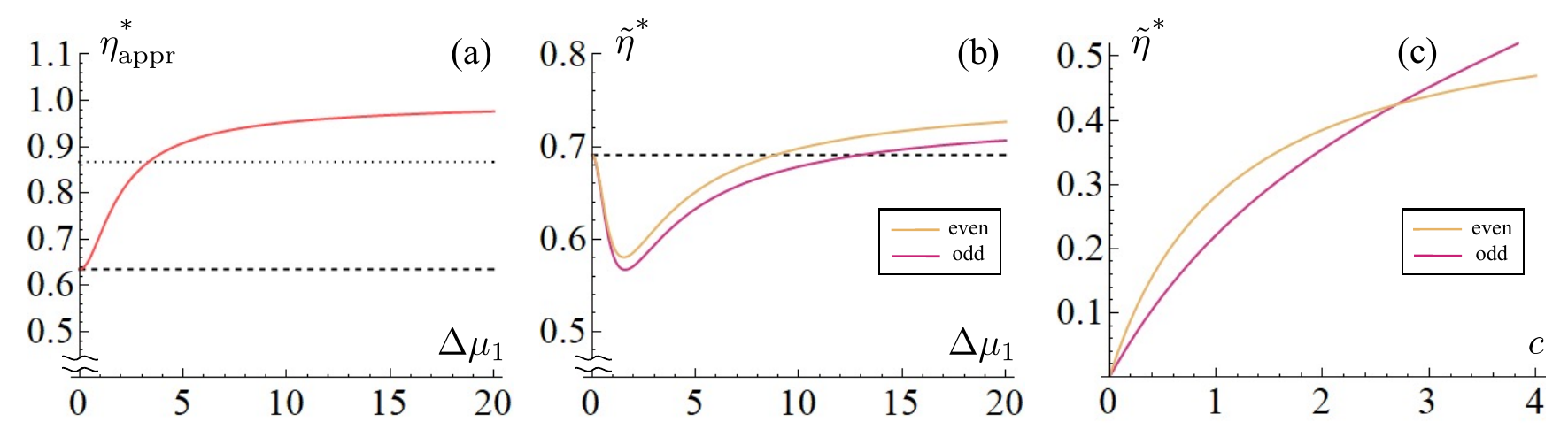}% Here is how to import EPS art
\caption{\label{fig:main}
The main results of this study.  
(a) The apparent EMP of the active heat engine (solid curve) and its passive counterpart (dashed line) compared with the Carnot efficiency (dotted line) for $\Delta\mu_2=0$, $c=3.75$, $\tau_1=3$, and $T_1=7.5$.
(b) The thermodynamically consistent EMPs of active heat engines with even/odd-parity self-propulsion compared with their passive counterpart for $\Delta\mu_2=0$, $c=3.75$, $\tau_1=2.375$, and $T_1=5$.
(c) The thermodynamically consistent EMPs with even/odd-parity self-propulsion for $\Delta\mu_1=1$, $\Delta\mu_2=0$, $\tau_1=0.875$ and $T_1=3$. For all figures of this paper, we use $K=2$ and $\Gamma = \gamma_1=\gamma_2=\gamma_1'=\gamma_2'=\zeta=\zeta'=T_2=1$ unless otherwise mentioned.
}
\end{figure*}

The rest of the paper is organized as follows. First, we briefly point to the main results of this paper in Sec.~\ref{sec:main}. Then, in Sec.~\ref{sec:chem_aoup}, we present minimal thermodynamically consistent models of a single AOUP driven by the constant chemical driving for both even-parity and odd-parity self-propulsion. We also clarify the energetics of the models and relate them to the entropy productions by the methods of stochastic thermodynamics. Based on these, in Sec.~\ref{sec:chem_AHE}, we propose a model for fuel-consuming active heat engines, which are simultaneously coupled to two reservoirs at different temperatures and operate in the steady state. The apparent dynamics of the engine is equivalent to the one studied in \cite{lee2020brownian}, but we assess the performance of the engine using a thermodynamically consistent definition of efficiency derived by stochastic thermodynamics. Then, in Sec.~\ref{sec:emp}, we apply our definition of efficiency to derive exact expressions for the efficiency at maximum power (EMP). We compare the EMPs of passive engines and active engines of both parities, deducing some design principles for active heat engines from the results. In Sec.~\ref{sec:tighterbound}, we show that the engines driven by the odd-parity propulsion force have a tighter upper bound on their efficiencies than given by the second law of thermodynamics. Finally, we summarize our findings and discuss possible future investigations in Sec.~\ref{sec:summary}.

\section{Main results}
\label{sec:main}

Before going into detail, we briefly point to the main findings of this study. Applying the theoretical approach described in \cite{dadhichi2018origins}, we consider two different thermodynamically consistent models of the AOUP that propels itself by consuming some chemical fuel. The first model features an even-parity self-propulsion force that does not change sign under time reversal. It can be regarded as describing an {\em active dimer} described by Eq.~\eqref{eq:model_even_x_2}. Meanwhile, the second model features an odd-parity self-propulsion force that changes sign under time reversal, which is described by Eq.~\eqref{eq:model_odd_p_2}. See Fig.~\ref{fig:schematic} for schematic illustrations of these two models. 

Applying the mechanochemical coupling used in these models, we study the efficiency of the fuel-driven active heat engine described by Eq.~\eqref{eq:gyrator}.

Using the standard methods of stochastic thermodynamics, we show that the entropy production of this engine always satisfies the Clausius relation stated in Eq.~\eqref{eq:clausius_ahe}, whose lower bound naturally leads to the expression for the thermodynamically consistent engine efficiency shown in Eq.~\eqref{eq:eta_comp}. This efficiency differs from the apparent engine efficiency considered in the previous studies~\cite{krishnamurthy2016micrometre,lee2020brownian}, stated in Eq.~\eqref{eq:eta_appr}.

When the maximum power is achieved, the apparent EMP $\eta_\mathrm{appr}^*$ is always higher for the active engine ($\Delta\mu_1 > 0$) than for its passive counterpart ($\Delta\mu_1 = 0$), see Eq.~\eqref{eq:eta_appr_a1} and Fig.~\ref{fig:main}(a). 

Meanwhile, using the definition of the engine efficiency we propose, the EMP of the active engine is greater than the passive counterpart only when the chemical driving $\Delta\mu_1$ is sufficiently strong, see Eq.~\eqref{eq:cEMP1} and Fig.~\ref{fig:main}(b). 

Finally, the even-parity (odd-parity) active engine achieves a higher EMP when the size scale of the engine (determined by the parameter $c$) is small (large) enough, see Eq.~\eqref{eq:evenminusodd} and Fig.~\ref{fig:main}(c). 

\section{Chemically driven AOUP}
\label{sec:chem_aoup}

The AOUP is one of the simplest models of the active particle dynamics. It assumes that the self-propulsion force of the active particle behaves like a noise whose autocorrelation decays exponentially in time, breaking the fluctuation-dissipation theorem (FDT). In one dimension, the dynamics of the AOUP is described by the following equation of motion:
\begin{align} \label{eq:aoup}
	\dot{X} = -\frac{1}{\Gamma} \, V'(X) + v + \xi_X.
\end{align}
Here $X$ denotes the position of the AOUP, $\Gamma$ the friction coefficient, $V(X)$ the external potential, $v$ the self-propulsion, and $\xi_X$ the thermal noise. The variables $v$ and $\xi_X$ are both Gaussian noises whose statistics satisfy
\begin{subequations} \label{eq:aoup_noise}
  \begin{align}
	\langle\xi_X(t)\rangle &= 0, &\langle\xi_X(t)\xi_X(t')\rangle &= \frac{2T}{\Gamma}\,\delta(t-t'),\label{eq:white}\\
	\langle v(t) \rangle &= 0, &\langle v(t)v(t') \rangle &= \frac{D_\mathrm{a}}{\tau}\mathrm{e}^{-|t-t'|/\tau},\label{eq:colored}
  \end{align}
\end{subequations}

where $T$ is the temperature, and $D_\mathrm{a}$ is the active contribution to the particle's diffusion coefficient. These relations indicate that $\xi_X$ is a {\em white} noise, while $v$ is a {\em colored} noise with a characteristic time scale $\tau$. The disagreement between this noise time scale $\tau$ and the instantaneous friction force $-\Gamma \dot{X}$ implied by Eq.~\eqref{eq:aoup} leads to the breaking of the FDT, thereby ensuring that the self-propulsion $v$ drives the particle out of equilibrium. However, the dynamics of $v$ is modeled only at the phenomenological level, so it is unclear from which dissipation forces the nonequilibrium driving of the system originates from.

In this section, we introduce a thermodynamically consistent, yet simple model of the AOUP which incorporates the constant chemical driving as the origin of the self-propulsion $v$. Towards this aim, we first present a general recipe for a system of Langevin equations that reaches equilibrium. Then we apply the recipe to construct the desired model for a single AOUP, whose energetics can be clearly identified.

\subsection{Recipe for an equilibrating Langevin system}

Our goal is to first construct a system of Langevin equations that reach equilibrium if there is no external driving, and then to add the external driving to keep the system {\em active}. Keeping this in mind, we consider an overdamped system described by the state vector $\mathbf{q}$, whose corresponding free energy is given by $F(\mathbf{q})$. In equilibrium, the system must satisfy the following conditions: (i) the steady-state distribution $p_\mathrm{s}$ must follow the Gibbs measure $p_\mathrm{s} \propto \mathrm{e}^{-F(\mathbf{q})/T}$, where $T$ is the temperature; (ii) the system satisfies the {\em detailed balance} (DB), or sometimes called {\em microreversiblity}, and thus becomes time-reversal symmetric in the steady state.

%% 230526 modification %%
%the {\em irreversible} probability current $\mathbf{J}^\mathrm{irr}(\mathbf{q}) \equiv \frac{1}{2}[\mathbf{J}(\mathbf{q})+\mathcal{E} \mathbf{J}(\mathcal{E}\mathbf{q})]$ must vanish in the steady state. Here $\mathcal{E}=\mathrm{diag}(\epsilon_{1},\ldots,\epsilon_{N})$ is the time-reversal operator with $\epsilon_i = +1$ ($-1$) if the $i$-th coordinate corresponds to an even-parity (odd-parity) variable. These conditions ensure that the system satisfies the {\em detailed balance} (DB) and thus becomes fully time-reversal symmetric in the steady state.

%For a given Hamiltonian $H(\mathbf{q})$ and finite temperature $T$, we aim to construct a Langevin system and the corresponding Fokker-Planck equation which relaxes to thermodynamic equilibrium at steady-state. This %guaranteed equilibration is enabled by assigning appropriate reciprocal coupling coefficients between the variables.

All the conditions listed above are satisfied by a system of Langevin equations (for $i = 1,\ldots,N$)
\begin{align}\label{general_recipe}
&\dot{q}_i=\sum_j \left[-(\Gamma_{ij}+R_{ij})\frac{\partial {F}}{\partial q_j} + T\frac{\partial}{\partial q_j}(\Gamma_{ij} + R_{ij})\right]+\xi_{i},\nonumber\\
&\big\langle\xi_i(t)\big\rangle =0,\quad
\big\langle\xi_i(t)\,\xi_j(t')\big\rangle=2T\Gamma_{ij}\,\delta(t-t'),
\end{align}
provided that the {\em dissipative} response coefficients $\Gamma_{ij}$ and the {\em reactive} response coefficients $R_{ij}$ satisfy the {\em Onsager reciprocal relations}~\cite{onsager1931reciprocali,onsager1931reciprocalii}
\begin{subequations}
\label{eq:onsager}
\begin{eqnarray}
\Gamma_{ij}(\mathbf{q})=\epsilon_i\epsilon_j \Gamma_{ij}(\mathcal{E}\mathbf{q})
=\Gamma_{ji}(\mathbf{q}),
\label{Onsager_even}
\\
R_{ij}(\mathbf{q})=-\epsilon_i\epsilon_j R_{ij}(\mathcal{E}\mathbf{q})=-R_{ji}(\mathbf{q}).
\label{Onsager_odd}
\end{eqnarray}
\end{subequations}
Here $\mathcal{E}=\mathrm{diag}(\epsilon_{1},\ldots,\epsilon_{N})$ is the time-reversal operator with $\epsilon_i = +1$ ($-1$) if the $i$-th coordinate corresponds to an even-parity (odd-parity) variable. We note that the flux $\dot{q}_i$ and the dissipative response $-\Gamma_{ij}\partial F/\partial q_j$ (reactive response $-R_{ij}\partial F/\partial q_j$) must have the opposite signs (same sign) under time reversal. The justification of this recipe is discussed in Appendix~\ref{apdx_generalrecipe}.

\subsection{Modeling the Fuel-Driven AOUP}

How do we apply the above recipe to construct a thermodynamically consistent model with the AOUP dynamics? We start by assuming that the self-propulsion $v$ is determined by the internal structure of the AOUP, which in itself follows the Ornstein-Uhlenbeck process
\begin{align} \label{eq:v_dyn}
	&\dot{v} = -\frac{1}{\tau} v + \xi_v,\nonumber\\
	&\big\langle \xi_v(t)\big\rangle = 0,
	\quad\big\langle\xi_v(t)\xi_v(t')\big\rangle = \frac{2D_\mathrm{a}}{\tau^2}\delta(t-t').
\end{align}
One can easily show that $v$ satisfying the above equations exhibits the exponentially decaying autocorrelation shown in Eq.~\eqref{eq:colored} in the steady state. But we are yet to decide which internal state determines $v$. We propose two different scenarios.

\subsubsection{Even-parity scenario}
We regard the AOUP as a dimer composed of two different species of monomers, see the dimer consisting of two distinguishable particles in Fig.~\ref{fig:schematic}(a). For the moment, we disregard the nonequilibrium driving on the dimer. Then we may write the free energy of the system composed of the dimer and the chemical fuel as follows:
\begin{align}
	F(X,x,n) = V(X)+\frac{1}{2}kx^2+f(n).
\end{align}
Here $x$ is the displacement of the brighter monomer from the dimer center, $k$ is the spring constant of the harmonic potential binding the monomers together, and $f(n)$ is the fuel contribution to the free energy, $n$ denoting the fuel concentration. Note that $X$, $x$, and $n$ all have even parities; thus, all fluxes (forces) are bound to have odd (even) parities, which means that the system can only have dissipative response coefficients.

Based on these considerations and the recipe shown above, we propose the following system of Langevin equations that reaches equilibrium:
\begin{subequations} \label{eq:model_even_x}
  \begin{align}
    \dot{X}
    &=-\frac{1}{\Gamma}V'(X)
    +\frac{\zeta x}{\Gamma}f'(n) + \xi_X,
    \label{eq:even_X_dyn}
    \\
    \dot{x}
    &=-\frac{k}{\gamma}x+\xi_{x},
    \label{eq:x_dyn}
    \\
    \dot{n}
    &=-\Gamma_{nn} \,f'(n)+\frac{\zeta x}{\Gamma} V'(X)+\xi_{n}.
    \label{eq:even_c_dyn}
  \end{align}
\end{subequations}
These originate from the choice of the response coefficients $\Gamma_{XX}=1/\Gamma$, $\Gamma_{Xn}=\Gamma_{nX}=-\zeta x/\Gamma$, $\Gamma_{xx}=1/\gamma$, where $\gamma$ and $\zeta$ are positive coefficients, with the other response coefficients being zero. These also imply that the noise components $\xi_X$, $\xi_x$, and $\xi_n$ all have zero means, and their correlations are given by
\begin{subequations} \label{eq:even_xi_corrs}
  \begin{align}
    \big\langle\xi_X(t)\xi_X(t')\big\rangle
    &= \frac{2T}{\Gamma}\,\delta(t-t'),
    \\
    \big\langle\xi_x(t)\xi_x(t')\big\rangle
    &= \frac{2T}{\gamma}\,\delta(t-t'),
    \\
  	\big\langle\xi_X(t)\xi_n(t')\big\rangle
    &= -2T\,\frac{\zeta x}{\Gamma}\,\delta(t-t'),\label{eq:even_xi_corrs_Xc}
    \\
    \big\langle\xi_n(t)\xi_n(t')\big\rangle
    &=2T\,\Gamma_{nn}\,\delta(t-t'), \label{eq:even_xi_corrs_cc}
  \end{align}
\end{subequations}
while the other correlations are all zero. We note that the noise correlation matrix $\mathbb{M}$, whose elements are given by $M_{ij}(t-t') \equiv \langle\xi_i(t)\xi_j(t')\rangle$, must be positive semi-definite. This implies 
\begin{align} \label{eq:Gamma_cc_ineq}
	\frac{\Gamma_{nn}}{\Gamma}-\left(\frac{\zeta x}{\Gamma}\right)^2 \ge 0, \quad
	\therefore \Gamma_{nn} \ge \frac{\zeta^2x^2}{\Gamma}.
\end{align}
We choose
\begin{align} \label{eq:Gamma_cc}
	\Gamma_{nn} = \frac{\zeta^2 x^2}{\Gamma},
\end{align}
so that we minimize the dissipation associated with the fuel dynamics while guaranteeing the inequality~\eqref{eq:Gamma_cc_ineq}. The consequence of this choice will be discussed again shortly. With all the response coefficients thus fixed, Eqs.~\eqref{eq:model_even_x}, \eqref{eq:even_xi_corrs}, and \eqref{eq:Gamma_cc} describe the equilibrium dynamics of a dimer coupled to the fuel at temperature $T$.

Now, to make the apparent dynamics of $X$ equivalent to the AOUP shown in Eqs.~\eqref{eq:aoup} and \eqref{eq:aoup_noise}, we fix $f'(n) = \Delta\mu$, which amounts to applying a constant chemical driving to the system. Also, Eqs.~\eqref{eq:even_xi_corrs_Xc}, \eqref{eq:even_xi_corrs_cc}, and \eqref{eq:Gamma_cc} imply that the noise $\xi_n$ can be rewritten as $\xi_n(t) = -\zeta x \xi_X(t)$. Taking all of these into account, the dynamics shown in Eq.~\eqref{eq:model_even_x} changes to
\begin{subequations} \label{eq:model_even_x_2}
  \begin{align}
    \dot{X}
    &=-\frac{1}{\Gamma}V'(X)
    +\frac{\zeta x}{\Gamma}\Delta\mu + \xi_X,
    \label{eq:even_X_dyn_2}
    \\
    \dot{x}
    &=-\frac{k}{\gamma}x+\xi_{x},
    \label{eq:x_dyn_2}
    \\
    \dot{n}
    &= -\zeta x \dot{X}.
    \label{eq:even_c_dyn_2}
  \end{align}
\end{subequations}
A comparison of this dynamics with Eqs.~\eqref{eq:aoup}, \eqref{eq:aoup_noise}, and \eqref{eq:v_dyn} shows that $X$ follows the AOUP dynamics with
\begin{align}
	v = \frac{\zeta x \Delta\mu}{\Gamma}, \quad \tau = \frac{\gamma}{k}, \quad D_\mathrm{a} = \left(\frac{\zeta\Delta\mu}{\Gamma k}\right)^2\gamma T.
\end{align}
Multiplying the friction coefficient $\Gamma$ to both sides of Eq.~\eqref{eq:even_X_dyn_2}, it is clear that $\zeta x \Delta\mu$ is the self-propulsion force acting on the AOUP. Thus, in this scenario, the parity of the self-propulsion is the same as that of $x$, which is an even-parity variable. Moreover, Eq.~\eqref{eq:even_c_dyn_2} shows that the fuel consumption is tightly coupled to the motion of the AOUP. This is the consequence of the choice of $\Gamma_{nn}$ made in Eq.~\eqref{eq:Gamma_cc}; in other words, the physical meaning of Eq.~\eqref{eq:Gamma_cc} is that the consumed fuel thoroughly contributes to the particle dynamics without any wasteful background reactions. When this assumption is not fulfilled, one can, for example, add a positive constant to $\Gamma_{nn}$. This introduces a constant background fuel consumption which does not directly contribute to the self-propulsion. One should expect such extra dissipation to be present in practical situations, such as the power consumption needed to keep the swimming bacteria alive and the extra heat dissipation from the light source that activates the self-propelled colloidal particles~\cite{malgaretti2021mechanical}. But here we focus on the fuel consumption directly relevant to the self-propulsion.

\subsubsection{Odd-parity scenario}

Now we consider case where the two monomers of the dimer are of the same type. The self-propulsion of this dimer does not come from the positions of the monomers, but from the rotation of the screw-like device attached between the monomers, see the dimer with a screw in the middle shown in Fig.~\ref{fig:schematic}(b). Denoting by $p$ the angular momentum of the screw, we may write the free energy of the system composed of the dimer and the chemical fuel as follows:
\begin{align}
	F(X,p,n) = V(X) + \frac{p^2}{2m} + f(n).
\end{align}
Here $m$ is the moment of inertia of the screw. In this scenario, our goal is to build a model where the self-propulsion changes sign under time reversal as does the flux $\dot{X}$. Such self-propulsion is bound to be a reactive response originating from $p$, which is the only odd-parity dynamical variable of the system.

Based on these considerations and the recipe shown above, we propose the following system of Langevin equations that reaches equilibrium:
\begin{subequations} \label{eq:model_odd_p}
  \begin{align}
  	\dot{X}
    &=-\frac{1}{\Gamma}V'(X)
    +\frac{\zeta' p}{\Gamma}f'(n) + \xi_X,
    \label{eq:odd_X_dyn}
    \\
    \dot{p}
    &=-\frac{\gamma'}{m}p+\xi_{p},
    \label{eq:p_dyn}
    \\
    \dot{n}
    &=-\Gamma_{nn} \,f'(n)-\frac{\zeta' p}{\Gamma} V'(X)+\xi_{c},
    \label{eq:odd_c_dyn}
  \end{align}	
\end{subequations}
These originate from the choice of the response coefficients $\Gamma_{XX}=1/\Gamma$, $R_{Xn}= -R_{nX}=-\zeta' p/\Gamma$, $\Gamma_{pp}=\gamma'$, where $\gamma'$ and $\zeta'$ are positive coefficients, with the other response coefficients being zero. These also imply that the noise components $\xi_X$, $\xi_p$, and $\xi_n$ all have zero means, and their correlations are given by
\begin{subequations} \label{eq:odd_xi_corrs}
  \begin{align}
    \big\langle\xi_X(t)\xi_X(t')\big\rangle
    &= \frac{2T}{\Gamma}\,\delta(t-t'),
    \\
    \big\langle\xi_p(t)\xi_p(t')\big\rangle
    &= 2T\,\gamma'\,\delta(t-t'),
    \\
    \big\langle\xi_n(t)\xi_n(t')\big\rangle
    &=2T\,\Gamma_{nn}\,\delta(t-t'), \label{eq:odd_xi_corrs_cc}
  \end{align}
\end{subequations}
while the other correlations are all zero. Again, the noise correlation matrix $\mathbb{M}$ must be positive semi-definite, which in this case requires $\Gamma_{nn} \ge 0$. As done in the even-parity scenario, we choose $\Gamma_{nn}$ so that the dissipation associated with the fuel dynamics is minimized. Thus we set $\Gamma_{nn} = 0$, {\em i.e.}, $\xi_n = 0$. Then, as in the previous case, we introduce a constant chemical driving by fixing $f'(n) = \Delta\mu$. These change the dynamics shown in Eq.~\eqref{eq:model_odd_p} to
\begin{subequations} \label{eq:model_odd_p_2}
  \begin{align}
  	\dot{X}
    &=-\frac{1}{\Gamma}V'(X)
    +\frac{\zeta' p}{\Gamma}\Delta\mu + \xi_X,
    \label{eq:odd_X_dyn_2}
    \\
    \dot{p}
    &=-\frac{\gamma'}{m}p+\xi_{p},
    \label{eq:p_dyn_2}
    \\
    \dot{n}
    &=-\frac{\zeta'p}{\Gamma}\left(-\Gamma\dot{X}+\zeta'p\Delta\mu+\Gamma\xi_X\right),
    \label{eq:odd_c_dyn_2}
  \end{align}	
\end{subequations}
A comparison of this dynamics with Eqs.~\eqref{eq:aoup}, \eqref{eq:aoup_noise}, and \eqref{eq:v_dyn} shows that $X$ follows the AOUP dynamics with
\begin{align}
	v = \frac{\zeta' p \Delta\mu}{\Gamma}, \quad \tau = \frac{m}{\gamma'}, \quad D_\mathrm{a} = \left(\frac{\zeta'\Delta\mu m}{\Gamma}\right)^2\frac{T}{\gamma'}.
\end{align}
Multiplying $\Gamma$ to both sides of Eq.~\eqref{eq:odd_X_dyn_2}, one can clearly see that $\zeta'p\Delta\mu$, an odd-parity term, plays the role of the self-propulsion force. We also note that Eq.~\eqref{eq:odd_c_dyn_2} only contains the fuel consumption associated with the screw rotation, without any background reaction that goes on even when $p = 0$. In this sense, the choice $\Gamma_{nn} = 0$ ensures the tight coupling between the fuel dynamics and the self-propulsion, as was done in the even-parity scenario.

\subsection{Energetics of the AOUP}
\label{ssec:aoup_energetics}

Now that we have fully modeled the dynamics of the chemically driven AOUP, we turn to the energetic interpretation of the model for each scenario.

\subsubsection{Even-parity scenario}
\label{sssec:even-parity}

So far, to derive Langevin equations with the proper mechanochemical coupling that ensures equilibration in the absence of driving, we have treated the fuel concentration $n$ as a dynamical variable of the system. However, with the chemical driving $\Delta\mu$ now fixed at a constant value, we regard the fuel supply as an external particle reservoir whose intensive properties do not change over time. In this viewpoint, now $X$ and $x$ are the only dynamical variables of the system, whose energy can be written as
\begin{align}
	E = V(X) + \frac{1}{2}kx^2.
\end{align}
Differentiating both sides with respect to time, we obtain
\begin{align}
	\dot{E}
	&= V'(X)\circ \dot{X} + kx \circ \dot{x} \nonumber\\
	&= \Gamma(-\dot{X} + \xi_X) \circ \dot{X} + \gamma(-\dot{x} + \xi_x) \circ \dot{x} + \zeta x \Delta\mu\,\dot{X},
\end{align}
where $\circ$ denotes the Stratonovich product~\cite{gardiner2009stochastic}, and the second equality is derived using Eqs.~\eqref{eq:even_X_dyn_2} and \eqref{eq:x_dyn_2}. Among the three terms on the rhs of the second equality, the last term is readily identified as the rate of {\em chemical work}
\begin{align} \label{eq:even_Wchem}
	\dot{W}_\mathrm{chem} \equiv -\Delta\mu \, \dot{n} = \zeta x \Delta\mu \, \dot{X},
\end{align}
where the second equality is found using Eq.~\eqref{eq:even_c_dyn_2}. Then, by the {\em first law of thermodynamics}, the rate of heat absorbed by the AOUP is
\begin{align} \label{eq:even_Q}
	\dot{Q}
	&= \dot{E}-\dot{W}_\mathrm{chem} \nonumber\\
	&= \Gamma(-\dot{X} + \xi_X) \circ \dot{X} + \gamma(-\dot{x} + \xi_x) \circ \dot{x}.
\end{align}
We note that this $\dot{Q}$ can be interpreted as the rate of work done on the AOUP by the reservoir forces $\Gamma(-\dot{X}+\xi_X)$ and $\gamma(-\dot{x}+\xi_x)$, which is in agreement with the standard microscopic definition of heat used in stochastic thermodynamics~\cite{sekimoto1997kinetic,sekimoto1998langevin,seifert2012stochastic}.

How do the work and heat identified above contribute to the dissipation of the system? To address this question, let us examine the conditional probability of the infinitesimal path
\begin{align} \label{eq:even_prob_forward}
	&\mathcal{P}[X+\dot{X}\,dt,x+\dot{x}\,dt,t+dt|X,x,t]\nonumber\\
	&\quad \sim \exp\Bigg[-\frac{\Gamma\,dt}{4T}\,\left(\dot{X}+\frac{1}{\Gamma}V'(X)-\frac{\zeta x \Delta\mu}{\Gamma}\right)^2\nonumber\\
	&\quad\quad\qquad - \frac{\gamma\,dt}{4T}\,\left(\dot{x}+\frac{kx}{\gamma}\right)^2+\frac{dt}{2}\big(V''(X)+k\big)\Bigg],
\end{align}
where $(X,\,x)$ in the above expression is to have the midpoint value between $(X,x)$ and $(X+\dot{X}dt,\,x+\dot{x}dt)$~\cite{wissel1979manifolds}. The conditional probability of the backward infinitesimal path can then be written as
\begin{align} \label{eq:even_prob_backward}
	&\mathcal{P}[X,x,t+dt|X+\dot{X}\,dt,x+\dot{x}\,dt,t]\nonumber\\
	&\quad \sim \exp\Bigg[-\frac{\Gamma\,dt}{4T}\,\left(-\dot{X}+\frac{1}{\Gamma}V'(X)-\frac{\zeta x \Delta\mu}{\Gamma}\right)^2\nonumber\\
	&~\quad\qquad - \frac{\gamma\,dt}{4T}\,\left(-\dot{x}+\frac{kx}{\gamma}\right)^2+\frac{dt}{2}\big(V''(X)+k\big)\Bigg].
\end{align}
According to the standard formalism of stochastic thermodynamics~\cite{seifert2012stochastic}, the environmental entropy production (EP) associated with the infinitesimal path is given by
\begin{align} \label{eq:even_Senv}
	dS_\mathrm{env} &\equiv \ln \frac{\mathcal{P}[X+\dot{X}\,dt,x+\dot{x}\,dt,t+dt|X,x,t]}{\mathcal{P}[X,x,t+dt|X+\dot{X}\,dt,x+\dot{x}\,dt,t]} \nonumber\\
	&= -\frac{dt}{T}\left[\Gamma(-\dot{X} + \xi_X) \circ \dot{X} + \gamma(-\dot{x} + \xi_x) \circ \dot{x}\right] \nonumber\\
	&= -\frac{dQ}{T},
\end{align}
where the last equality is obtained by comparison with Eq.~\eqref{eq:even_Q}. Thus, the heat identified in our model satisfies the Clausius relation for the EP. We note that there have been some previous proposals~\cite{puglisi2017clausius,marconi2017heat} of the Clausius relations for the EP of active particles, which rely on the notion of effective temperature of the {\em nonequilibrium bath} governing the active particle statistics. The Clausius relation shown in Eq.~\eqref{eq:even_Senv} differs from those in that it involves the temperature of and the heat exchange with the standard {\em equilibrium heat bath}.

\subsubsection{Odd-parity scenario}

Now we turn to the energetics of the odd-parity scenario. Again, we regard the fuel supply as an external particle reservoir, so the energy of the system can be written as
\begin{align}
	E = V(X) + \frac{p^2}{2m}.
\end{align}
Differentiating both sides with respect to time, we obtain
\begin{align}
	\dot{E}
	&= V'(X)\circ \dot{X} + \frac{p}{m}\circ \dot{p}	 \nonumber\\
	&= \left[-\Gamma \left(\dot{X}-\frac{\zeta' p\Delta\mu}{\Gamma}\right) +\Gamma \xi_X\right]\circ\left(\dot{X}-\frac{\zeta' p\Delta\mu}{\Gamma}\right)\nonumber\\
	&\quad + \left(-\frac{\gamma'}{m} p + \xi_p\right)\circ\frac{p}{m} \nonumber\\
	&\quad + \left[-\Gamma \left(\dot{X}-\frac{\zeta' p\Delta\mu}{\Gamma}\right) +\Gamma \xi_X\right]\frac{\zeta'p\Delta\mu}{\Gamma},
\end{align}
where the second equality is obtained by using Eqs.~\eqref{eq:odd_X_dyn_2} and \eqref{eq:p_dyn_2}. In a manner similar to the even-parity scenario, we identify the rates of chemical work $W_\mathrm{chem}$ and heat $Q$ absorbed by the AOUP as
\begin{align}
	\dot{W}_\mathrm{chem}
	&\equiv -\Delta\mu \,\dot{n}\nonumber\\
	&= 	\left[-\Gamma \left(\dot{X}-\frac{\zeta' p\Delta\mu}{\Gamma}\right) +\Gamma \xi_X\right]\frac{\zeta'p\Delta\mu}{\Gamma},\label{eq:odd_Wchem}\\
	\dot{Q} &= \left[-\Gamma \left(\dot{X}-\frac{\zeta' p\Delta\mu}{\Gamma}\right) +\Gamma \xi_X\right]\circ\left(\dot{X}-\frac{\zeta' p\Delta\mu}{\Gamma}\right)\nonumber\\
	&\quad + \left(-\frac{\gamma'}{m} p + \xi_p\right)\circ\frac{p}{m}, \label{eq:odd_Q}
\end{align}
so that the first law of thermodynamics $\dot{E} = \dot{Q} + \dot{W}_\mathrm{chem}$ is satisfied.

How can we interpret the expressions obtained above? Since the self-propulsion force $\zeta'p\Delta\mu$ changes sign under time reversal, the motion of the AOUP driven only by the force at velocity $\zeta'p\Delta\mu/\Gamma$ is in itself not an irreversible phenomenon, meaning such motion happens without dissipating any energy. Thus, the energy dissipation comes only from the {\em excess} velocity of the AOUP, $\dot{X} - \zeta'p\Delta\mu/\Gamma$; the frictional force, for the same reason, should be $\Gamma(-\dot{X} + \zeta'p\Delta\mu/\Gamma)$. This, together with the thermal force $\Gamma\xi_X$, forms the dissipative force applied by the thermal reservoir on the AOUP. Thus, Eq.~\eqref{eq:odd_Q} is a natural expression for the rate of energy dissipation at the reservoir. We note that similar expressions for heat in the presence of odd-parity self-propulsion were also proposed in \cite{speck2018active,dabelow2019irreversibility}.

Using stochastic thermodynamics, we can more explicitly check that $\dot{Q}$ identified in Eq.~\eqref{eq:odd_Q} quantifies the rate of energy dissipation. Towards this end, we examine the conditional probability of the infinitesimal path
\begin{align} \label{eq:odd_prob_forward}
	&\mathcal{P}[X+\dot{X}\,dt,p+\dot{p}\,dt,t+dt|X,p,t]\nonumber\\
	&~ \sim \exp\Bigg[-\frac{\Gamma\,dt}{4T}\,\left(\dot{X}+\frac{1}{\Gamma}V'(X)-\frac{\zeta' p \Delta\mu}{\Gamma}\right)^2\nonumber\\
	&\quad\quad~ - \frac{dt}{4\gamma' T}\,\left(\dot{p}+\frac{\gamma' p}{m}\right)^2+\frac{dt}{2}\bigg(V''(X)+\frac{1}{m}\bigg)\Bigg],
\end{align}
where $(X,\,p)$ in the above expression is to have the midpoint value between $(X,p)$ and $(X+\dot{X}dt,\,p+\dot{p}dt)$. The conditional probability of the backward infinitesimal path can then be written as
\begin{align} \label{eq:odd_prob_backward}
	&\mathcal{P}[X,-p,t+dt|X+\dot{X}\,dt,-p-\dot{p}\,dt,t]\nonumber\\
	&~ \sim \exp\Bigg[-\frac{\Gamma\,dt}{4T}\,\left(-\dot{X}+\frac{1}{\Gamma}V'(X)+\frac{\zeta' p \Delta\mu}{\Gamma}\right)^2\nonumber\\
	&~\quad\quad - \frac{dt}{4\gamma'T}\,\left(\dot{p}-\frac{\gamma' p}{m}\right)^2+\frac{dt}{2}\bigg(V''(X)+\frac{1}{m}\bigg)\Bigg],
\end{align}
where the odd parity of $p$ has been taken into account. Now, the environmental EP associated with the infinitesimal path is obtained as
\begin{align} \label{eq:odd_Senv}
	&dS_\mathrm{env} \equiv \ln \frac{\mathcal{P}[X+\dot{X}\,dt,p+\dot{p}\,dt,t+dt|X,p,t]}{\mathcal{P}[X,-p,t+dt|X+\dot{X}\,dt,-p-\dot{p}\,dt,t]} \nonumber\\
	&= -\frac{dt}{T}\left[-\Gamma \left(\dot{X}-\frac{\zeta' p\Delta\mu}{\Gamma}\right) +\Gamma \xi_X\right]\circ\left(\dot{X}-\frac{\zeta' p\Delta\mu}{\Gamma}\right) \nonumber\\
	&\quad -\frac{dt}{T}\left[\left(-\frac{\gamma'}{m}p+\xi_p\right)\circ\frac{p}{m}\right] 
	= -\frac{dQ}{T},
\end{align}
where the last equality, {\em i.e.}, the Clausius relation, comes from Eq.~\eqref{eq:odd_Q}. This confirms that $\dot{Q}$ identified in Eq.~\eqref{eq:odd_Q} indeed quantifies the rate of energy dissipation.

\section{Fuel-driven active heat engine}
\label{sec:chem_AHE}

So far, we have introduced a thermodynamically consistent model of a single fuel-consuming active particle following the AOUP. In order to describe how such a particle performs as a heat engine, we need to build a model which couples the particle to multiple thermal reservoirs. For this purpose, in this section, (i) we propose a fuel-consuming active heat engine operating between a pair of heat baths at different temperatures, (ii) clarify its energetics, and (iii) identify the engine efficiency bounded from above by the second law of thermodynamics.

\subsection{Modeling the fuel-driven active heat engine}

Our inspiration for the model comes from the {\em Brownian gyrator}, first proposed in \cite{filliger2007brownian} and implemented in \cite{chiang2017electrical}, which is a minimal model of a microscopic heat engine simultaneously coupled to two thermal reservoirs. Provided that the reservoirs are kept at different temperatures, the engine operates in a nonequilibrium steady state without any time-dependent protocol. A closely related variant, the linear Brownian engine, has also been studied and was shown to exhibit the Curzon-Ahlborn efficiency as the EMP, although it is not endoreversible~\cite{park2016efficiency}. Then the model was generalized to a heat engine coupled to two different active baths (or an active bath and an ordinary heat bath), with discussions of when the engine efficiency, defined at an apparent level, surpasses the Carnot efficiency~\cite{lee2020brownian}.

Employing a similar framework, we place our fuel-consuming engine in a two-dimensional space $(X_1,\,X_2)$. For each direction, the engine follows the fuel-consuming AOUP dynamics proposed in Sec.~\ref{sec:chem_aoup}, {\em i.e.}, Eqs.~\eqref{eq:model_even_x_2} or \eqref{eq:model_odd_p_2}. The constants characterizing each part of the dynamics, including the temperatures $T_1 \ge T_2$, may differ from each other; however, for simplicity, we assume that both $X_1$ and $X_2$ are constrained by the same harmonic potential $V(X_i) = (K/2)X_i^2$, and that both directions have the same friction coefficient $\Gamma$. To couple the dynamics of the two coordinates and extract work from the engine, we also apply a nonconservative force field ${\mathbf{f}}^{\textrm{nc}} = (\lambda_{\textrm{1}} X_2,\,\lambda_{\textrm{2}} X_1)$~\cite{park2016efficiency, lee2020brownian}. We note that, to be able to extract nonzero power from the engine in the steady state (where the internal energy of the engine saturates to a constant value), the force field does have to be nonconservative ($\lambda_1 \neq \lambda_2$).

Now, for the case where all variables have the even parity, the mechanical degrees of freedom $\mathbf{r} \equiv (X_1,\,X_2,\,x_1,\,x_2)$ obey the multivariate Langevin equation
\begin{align} \label{eq:gyrator}
    \dot{{\mathbf{r}}}
    =-\mathbb{K}
    \,{\mathbf{r}}
    +{\bm{\xi}
    }
\end{align}
with the $4 \times 4$ matrix
\begin{align} \label{eq:Kmat}
\mathbb{K}
=
\begin{bmatrix}
K/\Gamma & -\lambda_{\textrm{1}}/\Gamma & -\zeta_1\Delta\mu_1/\Gamma & 0 \\
-\lambda_{\textrm{2}}/\Gamma & K/\Gamma & 0 & -\zeta_2\Delta\mu_2/\Gamma \\
0 & 0 & 1/\tau_1 & 0 \\
0 & 0 & 0 & 1/\tau_2
\end{bmatrix}
\end{align}
and the noise satisfying $\langle\bm\xi(t)\rangle = 0$ and $\langle\bm{\xi}(t)\bm{\xi}^{\textrm{T}}(t')\big\rangle = 2\mathbb{D}\,\delta(t-t')$, where
\begin{align}
\mathbb{D} \equiv \mathrm{diag}\left(\frac{T_1}{\Gamma},\frac{T_2}{\Gamma}, \frac{T_1}{\gamma_1}, \frac{T_2}{\gamma_2}\right).
\end{align}
To change the dynamics of $X_i$ to that of an odd-parity AOUP, one can simply replace the variables and coefficients like $x_i \rightarrow p_i$, $k_i \rightarrow \frac{1}{m_i}$, $\gamma_i \rightarrow \frac{1}{\gamma_i'}$ and $\zeta_i \rightarrow \zeta_i'$. Meanwhile, the dynamics of the chemical degrees of freedom $c_1$ and $c_2$ have the same form as Eqs.~\eqref{eq:even_c_dyn_2} or \eqref{eq:odd_c_dyn_2} (after adding suitable subscript indices) depending on the parity of self-propulsion.

Provided that $\mathbf{f}_\mathrm{nc}$ satisfies $\lambda_1 \lambda_2 < K^2$, the mechanical degrees of freedom are stable and converge to a unique steady state that can be analytically obtained. Since Eq.~\eqref{eq:gyrator} simply defines a multi-dimensional Ornstein-Uhlenbeck process, $\mathbf{r}$ exhibits Gaussian statistics with zero mean in the steady state, so calculating its second moments fully determines the distribution. Calculations of those second moments are detailed in Appendix~\ref{apdx_covmtx}. We also note that, when $\mathbf{r}$ attains the steady state, the fuel concentrations $c_1$ and $c_2$ keep changing at constant rates, which can be calculated from the steady-state solution.

\subsection{Energetics of the fuel-driven active heat engine}

\begin{figure}[b]
\includegraphics{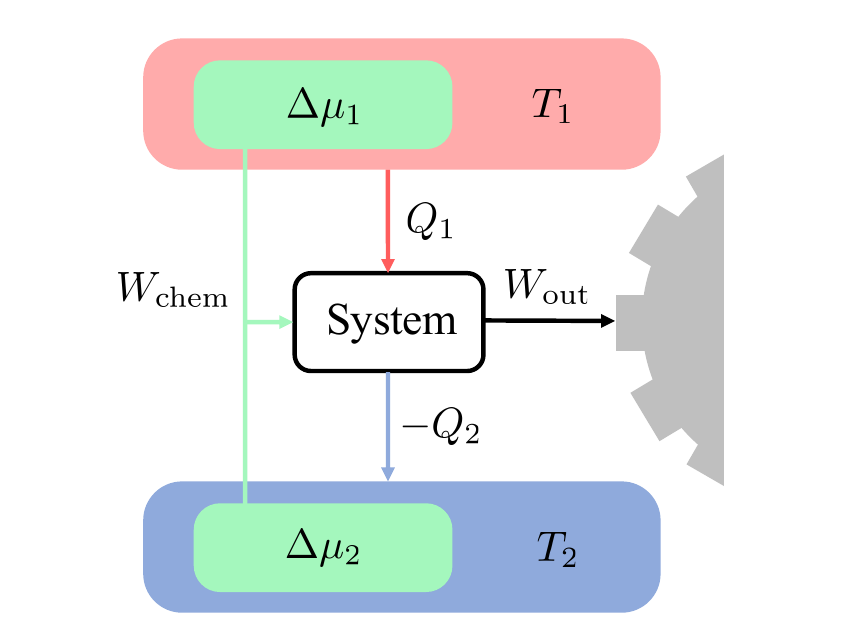}% Here is how to import EPS art
\caption{\label{fig:schematic_AHE}
The energetics of the fuel-driven active heat engine.
}
\end{figure}

Following the logic discussed in Sec.~\ref{ssec:aoup_energetics}, we can identify the work and the heat components of the energy flows generated by the active heat engine we have defined above. First of all, the work $W_{\mathrm{out},i}$ extracted by the $i$-th component of the external force $\mathbf{f}_\mathrm{nc}$ satisfies
\begin{align}
	\label{eq:wout_stoc}
	\dot{W}_{\mathrm{out},1} = -\lambda_1 X_2 \dot{X}_1, \quad
	\dot{W}_{\mathrm{out},2} = -\lambda_2 X_1 \dot{X}_2,
\end{align}
with the total extracted work given by $W_\mathrm{out} = W_{\mathrm{out},1}+W_{\mathrm{out},2}$. Meanwhile, the chemical work done along the $i$-th component can be written as
\begin{align} \label{eq:Wchem_ahe}
\dot{W}_{\mathrm{chem},i}=-\Delta\mu_i\,\dot{n}_i \quad \textrm{for $i = 1,\,2$,}
\end{align}
with the total chemical work $W_\mathrm{chem} = W_{\mathrm{chem},1}+W_{\mathrm{chem},2}$. Now, using the above and the first law of thermodynamics, the rate of heat absorbed by the engine through the $i$-th component is obtained as
\begin{align} \label{eq:first_law}
	\dot{Q}_i = \dot{E}_i + \dot{W}_{\mathrm{out},i}-\dot{W}_{\mathrm{chem},i},
\end{align}
where $E_i$ is the mechanical energy associated with the $i$-th component. These imply that $\dot{W}_{\mathrm{chem},i}$ and $\dot{Q}_i$ are respectively given by equations of the same form as Eqs.~\eqref{eq:even_Wchem} and \eqref{eq:even_Q} for the even-parity case and Eqs.~\eqref{eq:odd_Wchem} and \eqref{eq:odd_Q} for the odd-parity case, with matching subscript indices added. Then the rate of total heat absorbed by the engine is finally obtained as $\dot{Q} = \dot{Q}_1 + \dot{Q}_2$. The balance between all energy flows in the active heat engine identified above are schematically illustrated in Fig.~\ref{fig:schematic_AHE}. 

\subsection{Thermodynamically consistent engine efficiency}
\label{ssec:eta_comp}

Now we discuss how the energy flows identified above are bounded by the second law of thermodynamics. As was done in Sec.~\ref{ssec:aoup_energetics}, we can use the stochastic thermodynamics to relate the energy flows to the EP. More explicitly, in a manner analogous to Eqs.~\eqref{eq:even_Senv} and \eqref{eq:odd_Senv}, we identify the rate of environmental EP
\begin{align}
	dS_\mathrm{env} &\equiv \ln \frac{\mathcal{P}[\mathbf{r}+\dot{\mathbf{r}}\,dt,\,t+dt|\mathbf{r},\,t]}{\mathcal{P}[\mathcal{E}\mathbf{r},\,t+dt|\mathcal{E}(\mathbf{r}+\dot{\mathbf{r}}\,dt),\,t]}.
\end{align}
Then, after some algebra analogous to Eqs.~\eqref{eq:even_prob_forward}, \eqref{eq:even_prob_backward}, \eqref{eq:even_Senv}, \eqref{eq:odd_prob_forward}, \eqref{eq:odd_prob_backward}, and \eqref{eq:odd_Senv}, we can easily obtain the Clausius formula
\begin{align} \label{eq:clausius_ahe}
	\dot{S}_{\textrm{env}} =-\frac{1}{T_1}\dot{Q}_1-\frac{1}{T_2}\dot{Q}_2.
\end{align}
The rate of total EP is obtained by adding the rate of change of the Shannon entropy of the engine~\cite{seifert2012stochastic}. But, as long as we focus on the steady state, the average Shannon entropy of the engine stays constant. In that case, the total EP is on average equal to the environmental EP. Then, the Integral Fluctuation Theorem (IFT) implies the second-law inequality $\langle \dot{S}_\mathrm{env}\rangle \ge 0$, where $\langle \cdot \rangle$ denotes the average with respect to the steady-state distribution.

Using Eqs.~\eqref{eq:first_law} and \eqref{eq:clausius_ahe} in this inequality, we obtain
\begin{align} \label{eq:ineq0}
&\bigg\langle-\frac{\dot{Q}_1}{T_1}-\frac{\dot{Q}_2}{T_2}\bigg\rangle
=\bigg\langle-\frac{\dot{Q}_1}{T_1}-\frac{\dot{W}_{\textrm{out}}-\dot{Q}_1-\dot{W}_{\textrm{chem}}}{T_2}\bigg\rangle \nonumber\\
&\quad =-\frac{1}{T_2}\bigg[
\langle\dot{W}_{\textrm{out}}\rangle-\eta_\mathrm{C}\langle\dot{Q}_1\rangle-\langle\dot{W}_{\textrm{chem}}
\rangle\bigg]\geq 0,
\end{align}
where
\begin{align} \label{eq:eta_C}
	\eta_\mathrm{C} \equiv 1-\frac{T_2}{T_1}
\end{align}
is the Carnot efficiency. We note that
\begin{align} \label{eq:ineq1}
	\eta_\mathrm{C}\langle\dot{Q}_1\rangle + \langle \dot{W}_\mathrm{chem}\rangle = \langle\dot{W}_\mathrm{out}\rangle + T_2\langle\dot{S}_\mathrm{env}\rangle \ge \langle \dot{W}_\mathrm{out} \rangle,
\end{align}
so as long as the engine does operate as an engine ($\langle\dot{W}_\mathrm{out}\rangle > 0$), we have $\eta_\mathrm{C}\langle\dot{Q}_1\rangle + \langle \dot{W}_\mathrm{chem}\rangle > 0$~\cite{footnote}. Using this fact in Eq.~\eqref{eq:ineq0}, we identify a measure of engine performance bounded from above by the second law of thermodynamics: 
\begin{align} \label{eq:eta_comp}
\tilde{\eta}
\equiv\frac{\langle\dot{W}_{\textrm{out}}\rangle}{\eta_{\textrm{C}}\langle\dot{Q}_1\rangle+\langle\dot{W}_{\textrm{chem}}\rangle}\leq 1.
\end{align}
This measure, henceforth simply referred to as {\em efficiency}, naturally incorporates both heat injection and fuel consumption components of energy flows. Thus, this quantity is properly grounded upon the entire picture of how the engine converts the supplied energy into useful work.

We stress that the upper bound on the efficiency $\tilde{\eta}$ in Eq.~\eqref{eq:eta_comp} has been derived using only the laws of thermodynamics and two assumptions: (i) the engine extracts positive work in the steady state and (ii) the entropy is produced via contact with a pair of thermal reservoirs and satisfies the Clausius relation. Given these, the inequality is valid regardless of the details of the engine. Moreover, if the engine extracts work in a periodic state driven by a cyclic protocol, we can simply replace the heat and work rates in $\tilde{\eta}$ with corresponding quantities accumulated over a single period and still get the same inequality. Thus the above definition of $\tilde{\eta}$ is generally applicable to a broad range of engines.

Some comparisons of $\tilde{\eta}$ with the discussions of efficiency in the literature are in order. When there is no chemical driving ($\Delta\mu = 0$), Eq.~\eqref{eq:eta_comp} simply reduces to the Carnot upper bound on the efficiency of a heat engine. On the other hand, if the temperatures of the two baths are equal ($\eta_\mathrm{C} = 0$), then $\tilde{\eta} = \langle \dot{W}_\mathrm{out} \rangle/\langle \dot{W}_\mathrm{chem} \rangle \le 1$. This definition of efficiency has been used for isothermal active engines~\cite{pietzonka2019autonomous}, including molecular motors~\cite{parmeggiani1999energy,schmiedl2008efficiency,pietzonka2016universal}. Thus $\tilde{\eta}$ defined in Eq.~\eqref{eq:eta_comp} interpolates between the definitions of efficiency for the ordinary heat engines and the molecular motors. More recently, a similar measure of efficiency has been proposed in \cite{datta2022second}, which in place of $W_\mathrm{chem}$ uses an information-theoretical quantity describing the change of the probability distribution of the engine due to the nonzero $\Delta\mu$. In the presence of wasteful background chemical reactions that adds a large positive constant to $\Gamma_{nn}$ in Eq.~\eqref{eq:Gamma_cc}, $W_\mathrm{chem}$ would also increase by a large constant unrelated to the dynamics of the engine, trivially reducing $\tilde{\eta}$ far below $1$. Then the upper bound set by Eq.~\eqref{eq:eta_comp} would become a very loose bound, which is not useful for characterizing the performance of the engine. In such cases, the efficiency definition proposed by \cite{datta2022second} would be more useful as it focuses on the part of $W_\mathrm{chem}$ that affects the engine dynamics. But as long as the fuel consumption and the self-propulsion are tightly coupled, our definition of $\tilde{\eta}$ is similarly useful.

\subsection{Thermodynamics of the apparent efficiency}
\label{ssec:eta_appr}

Now that we have a clear energetic picture of the active heat engine, it is natural to ask how the previously defined notion of the apparent efficiency~\cite{krishnamurthy2016micrometre,lee2020brownian} exceeds the Carnot efficiency $\eta_\mathrm{C}$ without breaking the second law. We observe that, using our framework, the apparent efficiency discussed in the previous literature can be recast in a thermodynamically consistent manner as follows:
\begin{align}
    \label{eq:eta_appr}
    \eta_{\textrm{appr}} \equiv \frac{
    \langle\dot{W}_\mathrm{out}\rangle
    }
    {
    \langle\dot{Q}_{\textrm{1}}\rangle
    +
    \langle\dot{W}_{\textrm{chem,1}}\rangle
    }.
\end{align}
We note that this quantity is always equal to $1-\lambda_2/\lambda_1$ in our model, as derived in \cite{lee2020brownian}. In this definition, the denominator corresponds to the rate of energy injection from both thermal and fuel reservoirs to the position coordinate $X_1$. This definition might be a practical choice when we are unable to distinguish the thermal and the chemical parts of the injected energy. For example, in \cite{krishnamurthy2016micrometre,lee2020brownian}, the AOUP arises from the active bath without any accessible information about the fuel dynamics, so $\dot{W}_{\mathrm{chem},1}$ cannot be separated from $\dot{Q}_1$.

 An upper bound on $\eta_\mathrm{appr}$ is easily obtained using the inequality stated in Eq.~\eqref{eq:ineq1}:
\begin{align} \label{supercarnotnecessary}
    \eta_{\textrm{appr}}\leq\eta_{\textrm{C}}+\frac{(1-\eta_{\textrm{C}})\langle\dot{W}_{\textrm{chem,1}}\rangle+\langle\dot{W}_{\textrm{chem,2}}\rangle}{\langle\dot{Q}_1\rangle+\langle\dot{W}_{\textrm{chem,1}}\rangle}.
\end{align}
This inequality clearly shows that the positive chemical works ($\langle\dot{W}_{\mathrm{chem},i}\rangle > 0$) extends the thermodynamically allowed range of $\eta_\mathrm{appr}$ beyond the Carnot efficiency.

Meanwhile, after some manipulations, $\eta_\mathrm{appr}>\eta_\mathrm{C}$ leads to
\begin{align} \label{supercarnoteqv}
-\frac{\langle\dot{Q}_1\rangle+\langle\dot{W}_{\textrm{chem,1}}\rangle}{T_1}-\frac{\langle\dot{Q}_2\rangle+\langle\dot{W}_{\textrm{chem,2}}\rangle}{T_2} < 0.
\end{align}
If the chemical works are all zero, this condition violates the second law of thermodynamics expressed in Eq.~\eqref{eq:ineq0}; the apparent super-Carnot behavior $\eta_\mathrm{appr} > \eta_\mathrm{C}$ requires the presence of positive chemical works.

\section{Efficiency at maximum power}
\label{sec:emp}

Optimizing the design an engine is of theoretical and practical interest, but the aim of optimization should first be clarified. In this regard, the notion of EMP has been studied extensively due to the following reasons. First, it quantifies the efficiency achieved by an engine when it is most ``useful''. Second, some universal results regarding the EMP have been reported for a broad range of ordinary heat engines, especially the Curzon-Ahlborn efficiency $\eta_\mathrm{CA} \equiv 1 - \sqrt{T_2/T_1}$~\cite{park2016efficiency, curzon1975efficiency, cleuren2015universality, esposito2009universality}. 

%The \textit{dream engine} \cite{benenti2020power, lee2020exactly} is a hypothetical engine which simultaneously achieves maximum efficiency and finite power. However, it has been widely observed that there is a trade-off relation between the efficiency and power, e.g. quasi-static operation of the engine's protocol leads to vanishing power.

%For Langevin systems, Dechant and Sasa \cite{dechant2018entropic} argued that the magnitude of any current-like observables are bounded by EP from above. They apply it to microscopic heat engines, rigorously showing that they are impossible to reach a dream engine. Similarly applying that \textit{entropic bound} for our model, one can prove that the maximum efficiency is not attainable with finite power, unless the second moments of the dynamics diverge (not shown).

For the convenience of analysis, we change the external force parameters $\lambda_1$ and $\lambda_2$ to $r\equiv\lambda_1/\lambda_2$ and $c \equiv \lambda_1\lambda_2$. Then, for reasons to be clarified below, we search for $r$ maximizing the power for a fixed value of $c$. If only one of the two variables $X_1$ and $X_2$ is driven by the fuel, the condition for the maximum power can be analytically obtained with ease. In this section, we only present the results and compare the EMPs achieved by the passive heat engine and the active heat engine with even and odd-parity self-propulsion. For detailed derivations, see Appendix~\ref{apdx_explform}.

\subsection{Case $\Delta\mu_1>0$, $\Delta\mu_2=0$}

\subsubsection{Active vs. passive heat engines}

We first consider the case with $\Delta\mu_1>0$ and  $\Delta\mu_2=0$. For a fixed value of $c$, the power is maximized at $r=r^*$ with
\begin{align}
  r^* = \frac{a_1}{\sqrt{1-\eta_{\textrm{C}}}},
\end{align}
where 
\begin{align}
  a_1 \equiv \sqrt{1+\frac{\Gamma\tau_1^2\,\zeta_1^2\,\Delta\mu_1^2}{\gamma_1[(\Gamma + K\tau_1)^2-\tau_1^2 c]}}
\end{align}
for the even-parity case, and the corresponding expression for the odd-parity case can be obtained by the mapping $\gamma_1 \to 1/\gamma_1'$ and $\zeta_1 \to \zeta_1'$. Then the value of the maximum power (MP) is
\begin{align}
P^* = \left.\langle \dot{W}_\mathrm{out}\rangle\right|_{r=r^*} = \frac{T_2\,c}{2\Gamma K}\Big(\frac{a_1}{\sqrt{1-\eta_{\textrm{C}}}}-1\Big)^2.
\end{align}

Since $\eta_\mathrm{appr} = 1-\lambda_2/\lambda_1 = 1-1/r$, the apparent EMP is given by
\begin{align} \label{eq:eta_appr_a1}
\eta^*_\mathrm{appr} = 1-\frac{1}{a_1}\sqrt{1-\eta_{\textrm{C}}},
\end{align}
which reduces to the Curzon-Ahlborn efficiency in the passive limit $\Delta\mu_1 \to 0$, which corresponds to $a_1 \to 1$. We observe that both the MP and the apparent EMP monotonically increase as functions of $a_1$, which in turn monotonically increases with $\Delta\mu_1$. Thus, both the MP and the apparent EMP of the active heat engine are larger than their passive counterparts. As shown in Fig.~\ref{fig:main}(a), this apparent EMP can even surpass the Carnot efficiency $\eta_\mathrm{C}$.

Meanwhile, the thermodynamically consistent EMP is obtained as
\begin{align} \label{eq:cEMP1}
    \tilde{\eta}^*=\left[\frac{\eta_{\textrm{C}}}{1-\frac{\sqrt{1-\eta_{\textrm{C}}}}{a_1}}+b_1
    \frac{a_1^2-1}{\big(\frac{a_1}{\sqrt{1-\eta_{\textrm{C}}}}-1\big)^2}\right]^{-1},
\end{align}
where
\begin{align} \label{eq:b1_even}
	b_1 = b_{\mathrm{even},1} \equiv \frac{2K(\Gamma+K\tau_1)}{\tau_1 c}
\end{align}
for the even-parity engine and
\begin{align} \label{eq:b1_odd}
	b_1 = b_{\mathrm{odd},1} \equiv \frac{2K[\Gamma K + (K^2-c)\tau_1]}{\Gamma c}
\end{align}
for the odd-parity engine. Taking $\Delta\mu_1 \to 0$ ({\em i.e.}, $a_1 \to 1$) in Eq.~\eqref{eq:cEMP1}, we obtain
\begin{align}
	\tilde{\eta}^* = \frac{1-\sqrt{1-\eta_\mathrm{C}}}{\eta_\mathrm{C}}
	= \frac{\eta_\mathrm{CA}}{\eta_\mathrm{C}},
\end{align}
which yields the Curzon-Ahlborn efficiency $\eta_\mathrm{CA}$ in agreement with \cite{lee2020brownian}.

\begin{figure}%[b]
\includegraphics{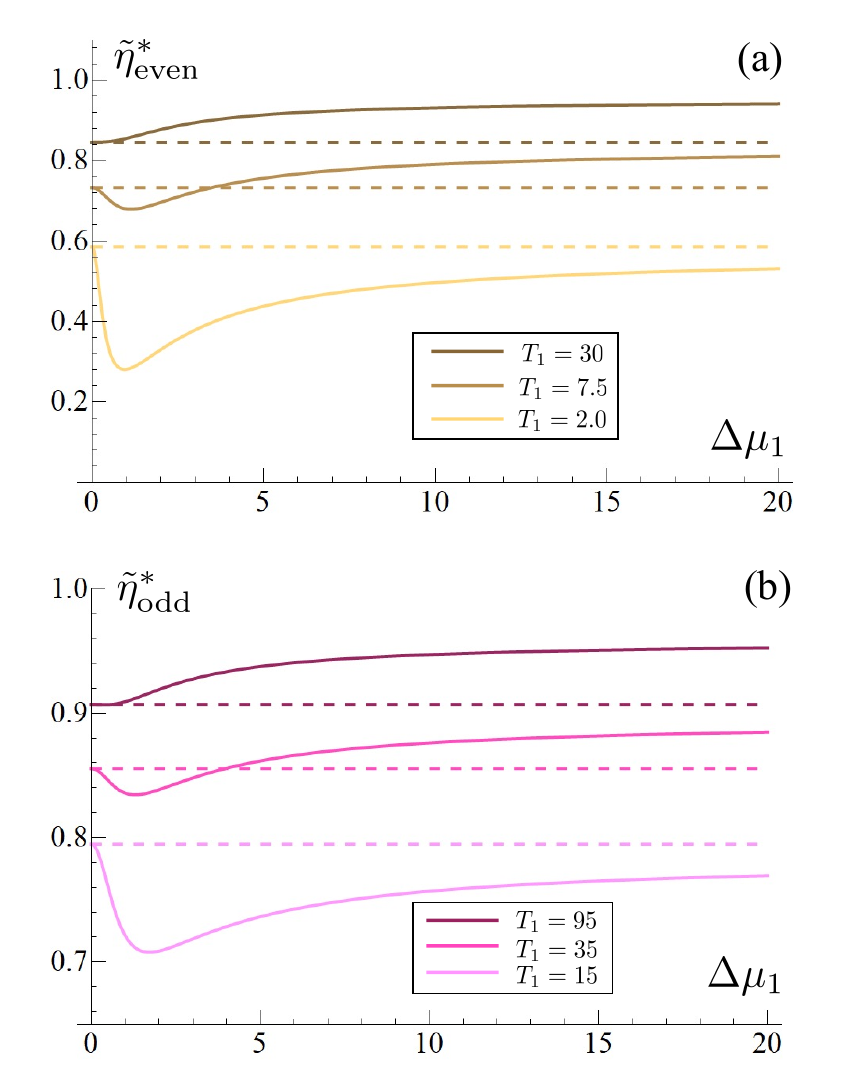}% Here is how to import EPS art
\caption{\label{fig:result_zdmuA} EMP of the active heat engine for various values of $T_1$ at fixed $T_2 = 1$ for (a) even-parity and (b) odd-parity self-propulsion. We used $c=3.5$, $\tau_1=5$, and $\Delta\mu_2=0$. Each dashed horizontal line indicates the passive EMP $\eta_\mathrm{CA}/\eta_\mathrm{C}$ at the corresponding value of $T_1$.}
\end{figure}

Now we discuss when the EMP of the active heat engine surpasses that of the passive counterpart. Comparing the passive EMP $\eta_\mathrm{CA}/\eta_\mathrm{C}$ obtained above with $\tilde{\eta}^*$, we obtain an inequality involving a quadratic polynomial of $a_1$. Noting that $a_1 \ge 1$ and that $b_1 \ge 2$ due to the stability condition $c < K^2$, there are three possible scenarios.

First, when $\eta_{\textrm{C}} \le \frac{b_1(b_1-2)}{(b_1-1)^2}$ (small temperature difference), the active EMP cannot surpass the passive EMP for any value of $\Delta\mu_1$, see the even-parity case with $T_1 = 2.0$ in Fig.~\ref{fig:result_zdmuA}(a) and the odd-parity case with $T_1 = 15$ in Fig.~\ref{fig:result_zdmuA}(b).

Second, when $\frac{b_1(b_1-2)}{(b_1-1)^2} < \eta_{\textrm{C}} < \frac{2b_1}{\sqrt{b_1^2+1}+b_1}$ (intermediate temperature difference), the active EMP is smaller than the passive EMP for small but positive $\Delta\mu_1$. However, the former eventually surpasses the latter as $\Delta\mu_1$ becomes larger, see the even-parity case with $T_1 = 7.5$ in Fig.~\ref{fig:result_zdmuA}(a) and the odd-parity case with $T_1 = 35$ in Fig.~\ref{fig:result_zdmuA}(b).

Third, when $\eta_{\textrm{C}} \ge \frac{2b_1}{\sqrt{b_1^2+1}+b_1}$ (large temperature difference), the active EMP is larger than the passive EMP for any positive $\Delta\mu_1$, see the even-parity case with $T_1 = 30$ in Fig.~\ref{fig:result_zdmuA}(a) and the odd-parity case with $T_1 = 95$ in Fig.~\ref{fig:result_zdmuA}(b).

As clearly shown in Fig.~\ref{fig:result_zdmuA}, the active EMP $\tilde{\eta}^*$ can exhibit a nonmonotonic dependence on the chemical driving $\Delta\mu_1$. This is an intriguing feature which illustrates that the behavior of a far-from-equilibrium system can be vastly different from a nonequilibrium system in the linear response regime.

\begin{figure*}%[b]
\includegraphics{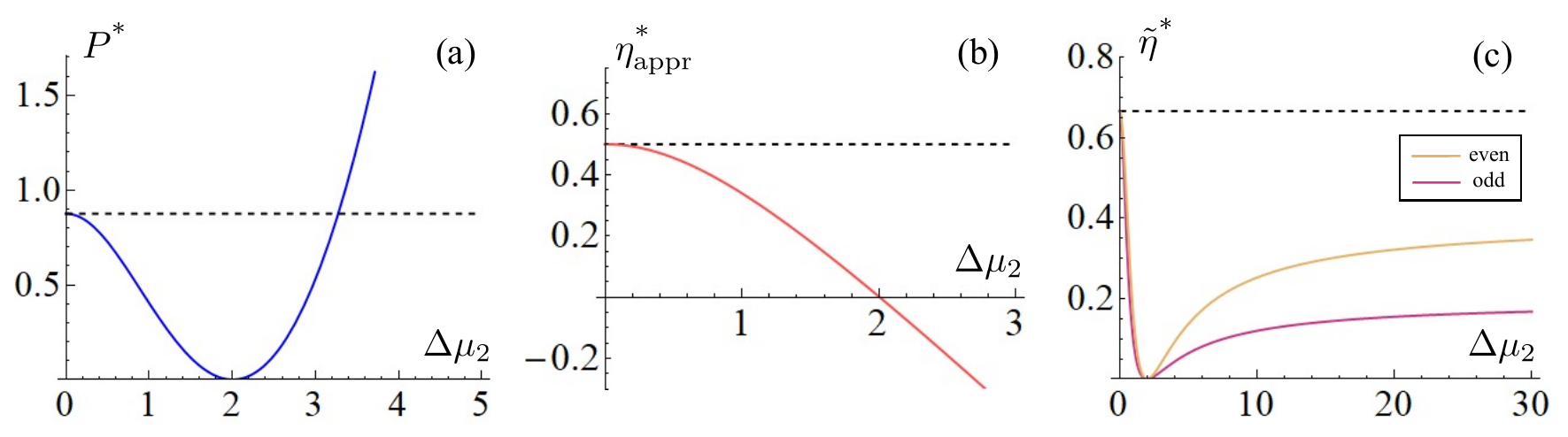}% Here is how to import EPS art
\caption{\label{fig:result_zdmuB}
Performance of the active heat engine when the chemical driving is attached to $X_2$. We use $c=3.5$, $\tau_2=5$, $T_1=4$ and $\Delta\mu_1=0$. (a) The MP, (b) the apparent EMP, and (c) the EMP are shown as functions of the chemical driving $\Delta\mu_2$ by solid curves. The dashed lines indicate the level of the corresponding quantity attained by the passive heat engine with $\Delta\mu_2 = 0$.}
\end{figure*}
\subsubsection{Even-parity vs. odd-parity engines}

From Eqs.~\eqref{eq:cEMP1}, \eqref{eq:b1_even}, and \eqref{eq:b1_odd}, we obtain
\begin{align}
    \label{eq:evenminusodd}
\frac{1}{\tilde{\eta}_{\mathrm{even}}^{*}}
- \frac{1}{\tilde{\eta}_{\mathrm{odd}}^{*}}
&=\textrm{(Positive constant)}\nonumber\\
&\quad\times\left[\Gamma^2 - (K^2-c)\tau_1^2\right],
\end{align}
so the sign of the lhs is solely determined by the dimensionless parameter
\begin{align}
	\alpha \equiv \frac{\Gamma}{\tau_1\sqrt{K^2-c}}.
\end{align}
The even-parity (odd-parity) engine achieves the higher EMP when $\alpha < 1$ ($\alpha > 1$). This is illustrated in Fig.~\ref{fig:main}(c) as the parameter $c$ is varied.

What is the physical significance of $\alpha$? Examining the exponential decays of the two-time correlation functions, we identify the three relaxation time scales:
\begin{align}
	\tau_+ \equiv \frac{\Gamma}{K + \sqrt{c}},\quad \tau_- \equiv \frac{\Gamma}{K - \sqrt{c}},\quad\tau_1.
\end{align}
Here $\tau_+$ and $\tau_-$ indicate the relaxation time scale of the AOUP within the spatial domain of the engine, while $\tau_1$ is the persistence time scale of the orientation of self-propulsion. Thus, given the typical velocity $v$ of the AOUP, the above time scales can be converted to the length scales $l_\pm \equiv v\tau_\pm$ reflecting the size of the engine and the length scale $l_1 \equiv v\tau_1$ corresponding to the persistence length. This shows that maximizing the power for a fixed $c$ amounts to optimizing the engine for a given size. Then, we can write
\begin{align}
	\alpha = \frac{\sqrt{l_+ l_-}}{l_1} = \frac{\textrm{(Length scale of the engine)}}{\textrm{(Persistence length)}},
\end{align}
which means that $\alpha$ quantifies the relative size of the engine with respect to the persistence length of the AOUP.

With this interpretation, we conclude that the even-parity (odd-parity) engine achieves the higher EMP when the engine is smaller (larger) than the persistence length of the AOUP. This can be intuitively understood in terms of the fuel consumption of the AOUP as follows. While the even-parity engine consumes fuel as the AOUP moves in space (see Eq.~\eqref{eq:even_c_dyn_2}), the odd-parity engine consumes fuel even when the AOUP does not move in space (see Eq.~\eqref{eq:odd_c_dyn_2}). When the engine is smaller than the persistence length, the AOUP tends to get stuck to the engine boundary with very slow actual motion in space. If this happens, the odd-parity engine spends much more fuel than the even-parity engine does, so the even-parity engine is more efficient. In contrast, when the engine is larger than the persistence length, the AOUP tends to move within the engine rapidly. If this motion is fast enough, the even-parity engine rapidly consumes fuel, whereas the fuel consumption rate of the odd-parity engine saturates (note that, for the single AOUP, Eq.~\eqref{eq:odd_c_dyn_2} can be rewritten as $\dot{n} = -\zeta'p V'(X)/\Gamma$, which does not explicitly involve $\dot{X}$). Thus, in this case, the odd-parity engine is more efficient than the even-parity engine.

\subsection{Case $\Delta\mu_1=0$, $\Delta\mu_2>0$}

Now we consider the case where the chemical driving is applied only to $X_2$ in contact with the cold thermal reservoir at temperature $T_2$. For a fixed $c$, the optimal value of $r$ is given by
\begin{align}
    r^*=\frac{1}{a_2\sqrt{1-\eta_{\text{C}}}},
\end{align}
where
\begin{align}
    a_2=\sqrt{1+\frac{\Gamma\,\tau_2^2\,\zeta_2^2\,\Delta\mu_2^2}{\gamma_2[(\Gamma + K\tau_2)^2-\tau_2^2 c]}}
\end{align}
for the even-parity case, and the corresponding expression for the odd-parity case can be obtained by the mapping $\gamma_2 \to 1/\gamma_2'$ and $\zeta_2\to\zeta_2'$. See Appendix~\ref{apdx_explform} for a detailed derivation. Using this result, the MP is obtained as
\begin{align}
	P^{\text{*}}=\frac{T_2\,c}{2K\Gamma}\left(\frac{1}{\sqrt{1-\eta_{\text{C}}}}-a_2\right)^2,
\end{align}
and the apparent EMP is given by
\begin{align}
	\eta_{\mathrm{appr}}^{*}=1-\frac{1}{r^*} = 1-a_2\sqrt{1-\eta_{\text{C}}}.
\end{align}
See Figs.~\ref{fig:result_zdmuB}(a) and \ref{fig:result_zdmuB}(b) for the behaviors of $P^{\text{*}}$ and $\eta_{\mathrm{appr}}^{*}$, respectively.

We may compare this expression with $\eta_\mathrm{appr}^*$ for the case $\Delta\mu_1 > 0$, $\Delta\mu_2 = 0$ shown in Eq.~\eqref{eq:eta_appr_a1}. Since both $a_1$ and $a_2$ cannot be less than $1$, $\eta_\mathrm{appr}^*$ obtained in Eq.~\eqref{eq:eta_appr_a1} is never less than $\eta_\mathrm{CA}$, while $\eta_\mathrm{appr}^*$ obtained above is never greater than $\eta_\mathrm{CA}$. Moreover, by increasing $a_2$ via increasing $\Delta\mu_2$, we see that the MP decreases to zero and then increases again, reaching the minimum for $a_2 = 1/\sqrt{1-\eta_\mathrm{C}}$, exactly where $\eta_\mathrm{appr}^*$ obtained above changes sign. This indicates that the denominator of $\eta_\mathrm{appr}$ in Eq.~\eqref{eq:eta_appr}, $\langle\dot{Q}_1\rangle$, becomes negative when MP is achieved for $a_2 > 1/\sqrt{1-\eta_\mathrm{C}}$ (note that $\langle \dot{W}_{\mathrm{chem},1}\rangle = 0$ here). In this regime, the chemical driving on $X_2$ is so strong that the engine operates by dissipating the heat even into the hot reservoir. Thus, here $\eta_\mathrm{appr}$ is a poor measure of the engine's efficiency.

In contrast, the thermodynamically consistent efficiency $\tilde{\eta}$ defined in Eq.~\eqref{eq:eta_comp} is still positive and bounded by $1$ as the positivity of its denominator is guaranteed by Eq.~\eqref{eq:ineq1}. The behavior of the EMP $\tilde{\eta}^*$ as $\Delta\mu_2$ is varied is shown in Fig.~\ref{fig:result_zdmuB}(c). 

The exact analytical form of the EMP is given by
\begin{align} \label{eq:cEMP2}
\tilde{\eta}^*
    =\left[\frac{\eta_{\text{C}}}{1-a_2\sqrt{1-\eta_{\text{C}}}}+b_2
    \frac{a_2^2-1}{\big(\frac{1}{\sqrt{1-\eta_{\text{C}}}}-a_2\big)^2}\right]^{-1},
\end{align}
where
\begin{align} \label{eq:b2_even}
	b_2 = b_{\mathrm{even},2} \equiv \frac{2K(\Gamma+K\tau_2)}{\tau_2 c}
\end{align}
for the even-parity engine and
\begin{align} \label{eq:b2_odd}
	b_2 = b_{\mathrm{odd},2} \equiv \frac{2K[\Gamma K + (K^2-c)\tau_2]}{\Gamma c}
\end{align}
for the odd-parity engine. Once again, we examine whether $\tilde{\eta}^*$ can be greater than its value in the passive limit $\eta_\mathrm{CA}/\eta_\mathrm{C}$. Noting that $\tilde{\eta}^* = \eta_\mathrm{CA}/\eta_\mathrm{C}$ yields a quadratic equation for $a_2$ and that $a_2 \ge 1$ and $b_2 \ge 2$ (due to the stability condition $K^2 > c$), we can show that $\tilde{\eta}^*$ cannot be greater than $\eta_\mathrm{CA}/\eta_\mathrm{C}$. 

As for the effects of the parity on the EMP, we obtain
\begin{align}
	\frac{1}{\tilde{\eta}_{\mathrm{even}}^{*}}
- \frac{1}{\tilde{\eta}_{\mathrm{odd}}^{*}}
&=\textrm{(Positive constant)}\nonumber\\
&\quad\times\left[\Gamma^2 - (K^2-c)\tau_2^2\right],
\end{align}
which is almost the same as Eq.~\eqref{eq:evenminusodd}, except for the replacement $\tau_1 \to \tau_2$. Thus the previous discussions of when the even-parity engine is more efficient than the odd-parity engine are also fully applicable in this case.

\section{Tighter bound on efficiency}\label{sec:tighterbound}
Thus far, we have defined and examined the efficiency $\tilde{\eta}$ of active heat engines, whose upper bound set by the second law of thermodynamics is $1$. But is there a tighter upper bound on $\tilde{\eta}$?

Indeed, stochastic thermodynamics has generalized the second law of thermodynamics, identifying different kinds of EP which are guaranteed to be nonnegative. Since we are dealing with an engine that operates in the steady state, the relevant type of EP is the {\em housekeeping} EP, which is associated with the maintenance of the nonequilibrium steady state. In the presence of odd-parity variables, the housekeeping EP can be further decomposed into two parts~\cite{lee2013fluctuation}: the part associated with the breaking of detailed balance in the steady state (whose rate is denoted by $\dot{S}_\mathrm{bDB}$) and the part associated with the breaking of the mirror symmetry of the steady-state distribution $p_{\textrm{s}}(\mathbf{r})=p_{\textrm{s}}(\mathcal{E}\mathbf{r})$ (whose rate is denoted by $\dot{S}_\mathrm{as}$). Only $\langle\dot{S}_\mathrm{bDB}\rangle \ge 0$ is guaranteed, which yields an inequality distinct from that shown in Eq.~\eqref{eq:ineq1}:
\begin{align}
\eta_\mathrm{C}\langle\dot{Q}_1\rangle + \langle \dot{W}_\mathrm{chem}\rangle 
&= \langle\dot{W}_\mathrm{out}\rangle + T_2\langle\dot{S}_\mathrm{bDB}\rangle + T_2\langle\dot{S}_\mathrm{as}\rangle \nonumber\\
&\ge \langle \dot{W}_\mathrm{out} \rangle + T_2\langle\dot{S}_\mathrm{as}\rangle.
\end{align}
Using the definition of $\tilde{\eta}$ in Eq.~\eqref{eq:eta_comp}, this inequality implies
\begin{align} \label{eq:eta_comp_tighterbound}
\tilde{\eta} \le 1 - \frac{T_2\langle\dot{S}_\mathrm{as}\rangle}{\eta_\mathrm{C}\langle\dot{Q}_1\rangle + \langle \dot{W}_\mathrm{chem}\rangle}
\end{align}
as long as $\langle\dot{W}_\mathrm{out}\rangle > 0$ (see the discussion below Eq.~\eqref{eq:ineq1}). This gives a tighter upper bound on $\tilde\eta$ if $\langle\dot{S}_\mathrm{as}\rangle > 0$.

\begin{figure}%[b]
\includegraphics{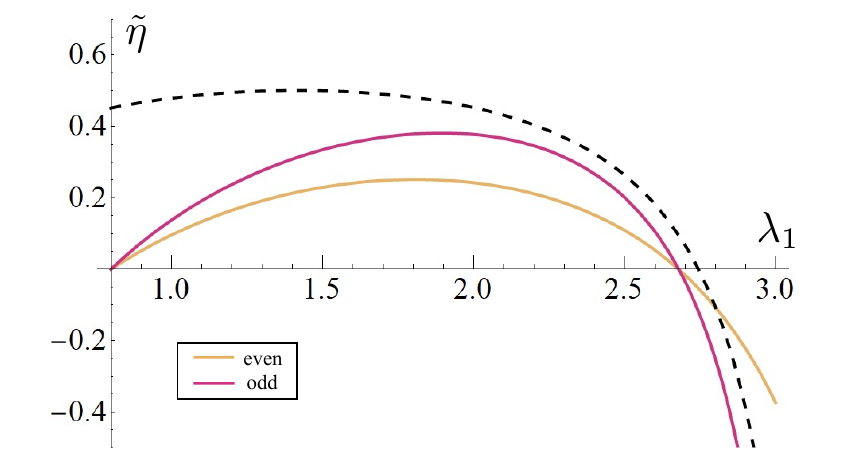}% Here is how to import EPS art
\caption{\label{fig:result_tighterbound} Tighter upper bound (black dashed curve) for odd-parity efficiency (dark solid curve). We used $\Delta\mu_1=1.75$, $\Delta\mu_2=0$, $\tau_1=0.3$, $T_1=3$ and $\lambda_2=0.8$. The efficiency of the engine with even-parity self propulsion (bright solid curve) is also shown for comparison.}
\end{figure}

When there are only even-parity variables, the mirror symmetry $p_{\textrm{s}}(\mathbf{r})=p_{\textrm{s}}(\mathcal{E}\mathbf{r})$ is trivially satisfied, so $\langle\dot{S}_\mathrm{as}\rangle = 0$. On the other hand, in the presence of odd-parity variables, the mirror symmetry is in general not guaranteed. Thus, here we focus on whether the above inequality yields a tighter upper bound on the efficiency of an odd-parity active heat engine.

For a diffusive system like the models considered here, there is an infinite number of possible definitions of $\dot{S}_\mathrm{bDB}$ and $\dot{S}_\mathrm{as}$, which can be parametrized by a single real number $\sigma$~\cite{yeo2016housekeeping}. As detailed in Appendix~\ref{apdx_bDB_as}, we follow their method to derive
\begin{align} \label{eq:Sas}
&\langle \dot{S}_\mathrm{as}\rangle =\sigma(1-\sigma)\nonumber\\
&\quad\qquad \times \left\langle\sum_{i=1}^2\left[\frac{T_i}{\Gamma}\left(\frac{\partial\phi^{\textrm{A}}}{\partial X_i}\right)^2
+\gamma_i'\, T_i\left(\frac{\partial\phi^{\textrm{A}}}{\partial p_i}\right)^2\right]\right\rangle,
\end{align}
where
\begin{align}
\phi^{\textrm{A}}(\mathbf{r})\equiv-\ln \,[p_\mathrm{s}(\mathbf{r})/p_\mathrm{s}(\mathcal{E}\mathbf{r})]
\end{align}
quantifies the extent to which the mirror symmetry is broken.

In Eq.~\eqref{eq:Sas}, the expression inside $\langle\cdot\rangle$ is always nonnegative, so $\langle\dot{S}_\mathrm{as}\rangle$ is maximized when $\sigma = 1/2$. Thus, in this case, Eq.~\eqref{eq:eta_comp_tighterbound} imposes a tighter upper bound on the efficiency $\tilde{\eta}$ of an odd-parity engine than the second law of thermodynamics does. This is illustrated in Fig.~\ref{fig:result_tighterbound} as the nonconservative force coefficient $\lambda_1$ is varied. The new upper bound imposed by Eq.~\eqref{eq:eta_comp_tighterbound} is significantly tighter than the original upper bound $\tilde\eta \le 1$. Also note that this new upper bound is applicable only to the odd-parity engine as exemplified by the efficiency of the even-parity engine surpassing the bound for large $\lambda_1$.

%Both EP components  that arises when the control parameters of the system change,   the canonical way of viewing general dynamical systems is to distinguish the transient and steady-state behaviors. Following this philosophy, the heat flows involving each behavior are identified \cite{oono1998steady,hatano2001steady}. From this, alternative decomposition of EP \cite{spinney2012entropy,seifert2012stochastic} unalike the traditional one $\Delta S_{\textrm{tot}}=\Delta S_{\textrm{sys}} + \Delta S_{\textrm{env}}$ is introduced. A part of EP resulting from the (continuous or abrupt) transiencies is identified as {\em excess} EP and denoted as $\Delta S_{\textrm{ex}}$. However, the remaining contribution $\Delta S_{\textrm{hk}}:=\Delta S_{\textrm{tot}}-\Delta S_{\textrm{ex}}$ is produced as a result of maintaining the system out of equilibrium, where the subscript $\textrm{hk}$ means {\em housekeeping}. Since our model already have reached the non-equilibrium steady state (NESS), the excess EP vanishes and the housekeeping EP is equal to $\Delta S_{\textrm{env}}$%(See Appendix \ref{apdx_bDB_as} for rigorous justification).

\section{Summary and outlook}
\label{sec:summary}

In this study, we proposed a thermodynamically consistent, analytically solvable model of the active heat engines based on the fuel-driven Active Ornstein-Uhlenbeck Process with either even-parity or odd-parity self-propulsion.

Our model, which stays active only due to the constant chemical driving, reflects how the fuel consumption dynamics should change depending on the self-propulsion parity. It also has a clear energetic interpretation for the entropy production in the form of the Clausius relation, which is lacking in the usual phenomenological models of active heat engines. This energetic picture allows us to define the efficiency of the engine as a ratio $\tilde{\eta}$ between two measurable energy fluxes, namely the extracted power and the energy flux arising from the thermal and the chemical driving forces. Moreover, the efficiency thus defined has an upper bound imposed by the second law of thermodynamics.

Taking this $\tilde{\eta}$ to be the proper measure of efficiency, we quantified the performance of the engine by examining its efficiency at maximum power $\tilde{\eta}^*$. First, we checked whether the {\em active} nature of the engine can make it more efficient than the passive engine in terms of $\tilde{\eta}^*$. Intriguingly, we found that $\tilde{\eta}^*$ may have a nonmonotonic dependence on the strength of the chemical driving, so that the active engine may become more efficient than the passive one when the chemical driving is strong enough. Second, we compared the performances of the even-parity and the odd-parity active engines. It turned out that the size of the engine matters: if the engine is larger (smaller) than persistence length of the particle, the odd-parity (even-parity) self-propulsion is more efficient. These results suggest interesting design principles that should be taken into account when constructing efficient and yet functioning active engines.

Finally, we explored the possibility of a tighter upper bound on $\tilde{\eta}$ than the one imposed by the second law of thermodynamics. Using the detailed structure of the housekeeping entropy production, we derived a tighter upper bound on the efficiency of the odd-parity engines and found an example where the bound is very close to the actual efficiency of the engine. 

These findings suggest various directions of future investigations. First, one may verify whether the design principles found in our study are indeed at work in more realistic examples of active heat engines, such as those made of Janus particles or swimmers propelled by a screw-like structure. Such systems typically involve hydrodynamic interactions with the liquid-like medium, making dynamic and thermodynamic descriptions much more challenging. Nonetheless, since we could explain the relative performance of the even-parity and the odd-parity engines using an intuitive argument based on a few length (or time) scales of the engine, we believe that the design principles will be generally applicable to more complicated systems.

Second, one may apply our model to a system of Active Ornstein-Uhlenbeck Particles and explore how their collective phenomena are related to the rate of energy dissipation in the entire system. Previous studies of such relations are based only upon the measure of apparent irreversibility~\cite{nemoto2019optimizing,fodor2016far,nardini2017entropy}, so while they address the question of whether a large-scale dissipative structure can also be achieved by an equilibrium system at the dynamical level, they do not address the question of how much fuel is required to maintain such structure. Our approach provides a useful framework for investigating the latter question for both even-parity and odd-parity active particles. We also note that a thermodynamically consistent framework for the energy dissipation that maintains a nonequilibrium structure at the field-theoretical level has been proposed in \cite{markovich2021thermodynamics}.

Third, one may explore the behaviors of our model under time-dependent protocols and quantify the performance of cyclic active heat engines. In particular, the question of a tighter bound on the engine efficiency becomes much more relevant in this case, because the {\em excess} entropy production, which is trivially zero in the steady state, also becomes a crucial part of the energy dissipation mechanism. Both the even-parity and the odd-parity engines will have a tighter bound on their efficiencies in this case. It would be also interesting to check whether thermodynamic uncertainty relations and speed limits on the entropy production yield interesting tradeoff relations involving the fuel consumption. The theoretical formalism for periodically driven engines in the linear response regime developed in \cite{brandner2015thermodynamics} might be useful for this purpose.

\begin{acknowledgments}
This work was supported by the National Research  Foundation of Korea Grant funded by the Korean Government (NRF-2020R1C1C1014436). YB also thanks Michael E. Cates, Patrick Pietzonka, Tomer Markovich, \'{E}tienne Fodor, Hyunggyu Park, and Jae Sung Lee for helpful discussions.
\end{acknowledgments}

\appendix

\section{Derivation of the recipe for an equilibrating Langevin system}
\label{apdx_generalrecipe}
Consider a Langevin system $\dot{\mathbf{q}}=\mathbf{A}(\mathbf{q})+\bm{\xi}$ with a state vector $\mathbf{q}=(q_1,\cdots,q_N)\in\mathbb{R}^N$ and the Gaussian white noise $\bm{\xi}$ satisfying $\langle
\bm\xi(t)\bm\xi(t')^\textrm{T}\rangle=2\mathbb{D}(\mathbf{q})\,\delta(t-t')$. Here the $N \times N$ diffusivity matrix $\mathbb{D}(\mathbf{q})=\{D_{ij}(\mathbf{q})\}$ is positive definite and symmetric, {\em i.e.}, $D_{ij}(\mathbf{q})=D_{ji}(\mathbf{q})$. The steady-state probability distribution $p_{\textrm{s}}(\mathbf{q})$ satisfies the Fokker-Planck equation (FPE)
\begin{align} \label{eq:FPE}
0=\frac{\partial p_\textrm{s}}{\partial t}=-\sum_i\frac{\partial}{\partial q_i}\left[A_i
%(\mathbf{r})
p_{\textrm{s}}
%(\mathbf{r})
-\sum_j\frac{\partial}{\partial q_j}\left(D_{ij}
%(\mathbf{r})
p_{\textrm{s}}
%(\mathbf{r})
\right)\right] \nonumber\\
=:-\sum_i \frac{\partial}{\partial q_i} J_i^{\textrm{s}}(\mathbf{q}),
\end{align}
where $J_i^\textrm{s}(\mathbf{q})$ is the steady-state probability current.

We assume the system to be in the linear response regime. Then the drift $\mathbf{A}(\mathbf{q})$ can be written as
\begin{align}
A_i(\mathbf{q})=-\sum_{j=1}^N \chi_{ij}(\mathbf{q})\frac{\partial F(\mathbf{q})}{\partial q_j}\,+\,\psi_i(\mathbf{q}),
	\label{drift_linear}
\end{align}
where $\chi_{ij}(\mathbf{q})$ is the response coefficient, and $F(\mathbf{q})$ is the free energy, so that $\partial F(\mathbf{q})/\partial q_j$ is the {\em thermodynamic force} in the $j$-direction. Meanwhile, $\psi_i(\mathbf{q})$ is a constant term independent of the thermodynamic forces, whose necessity will be clarified shortly.

We first require that the steady-state distribution of the system is given by the Gibbs measure
\begin{align}
	\label{equil_condition_1}
	p_{\textrm{s}}(\mathbf{q})\propto e^{-F(\mathbf{q})/T},
\end{align}
where $T$ is the temperature of the surrounding thermal reservoir. Applying this condition and Eq.~\eqref{drift_linear} to Eq.~\eqref{eq:FPE}, we obtain
\begin{widetext}
\begin{align}
	\label{FPE_expand}
	0=\sum_i p_\mathrm{s}\times\Bigg\{
	-\frac{\partial}{\partial q_i}\bigg(\psi_i-\sum_j\frac{\partial D_{ij}}{\partial q_j}\bigg)
	+\frac{1}{T}\bigg[\psi_i+\sum_j\bigg(T\frac{\partial\chi_{ji}}{\partial q_j}-2\frac{\partial D_{ij}}{\partial q_j}\bigg)\bigg]\frac{\partial F}{\partial q_i}
%\qquad \qquad \qquad	
	 \nonumber
	\\
	+\sum_j\bigg[\bigg(\chi_{ij}-\frac{1}{T}D_{ij}\bigg)\frac{\partial^2 F}{\partial q_i \partial q_j}-\frac{1}{T}\bigg(\chi_{ij}-\frac{1}{T}D_{ij}\bigg)\frac{\partial F}{\partial q_i}\frac{\partial F}{\partial q_j}
	\bigg]
	\Bigg\}.
\end{align}
\end{widetext}

Since this equation should hold for any $F(\mathbf{q})$, the coefficient of each $\partial^2 F/(\partial q_i\partial q_j)$ must be zero. That is,
\begin{align}
	\label{swap_sum_example}
\frac{1}{2}\sum_{i,j}\left[\left(\chi_{ij}-\frac{1}{T}D_{ij}\right)\frac{\partial^2 F}{\partial q_i \partial q_j} + \left(\chi_{ji}-\frac{1}{T}D_{ji}\right)\frac{\partial^2 F}{\partial q_j \partial q_i}\right]
\nonumber
\\
=\sum_{i,j}\left[\frac{1}{2}\left(\chi_{ij}+\chi_{ji}\right)-\frac{1}{T}D_{ij}\right]\frac{\partial^2 F}{\partial q_i \partial q_j}=0,
\end{align}
where the symmetry $D_{ij} = D_{ji}$ has been used. Now we introduce a decomposition $\chi_{ij}(\mathbf{q})=\Gamma_{ij}(\mathbf{q})+R_{ij}(\mathbf{q})$ so that
\begin{subequations}
\begin{align}
		\Gamma_{ij}(\mathbf{q}):=\frac{1}{2}\Big[\chi_{ij}(\mathbf{q})+\chi_{ji}(\mathbf{q})\Big],
		\\
		R_{ij}(\mathbf{q}):=\frac{1}{2}\Big[\chi_{ij}(\mathbf{q})-\chi_{ji}(\mathbf{q})\Big].
\end{align}	
\end{subequations}
Here $\Gamma_{ij}(\mathbf{q})$ captures the \textit{symmetric} part of the response coefficient, while $R_{ij}(\mathbf{q})$ is the \textit{antisymmetric} part, \textit{i.e.,} $R_{ij}(\mathbf{q})=-R_{ji}(\mathbf{q})$. According to Eq.~\eqref{swap_sum_example} and the symmetry of $\mathbb{D}$, the following should hold:
\begin{subequations}
\label{Gamma_conditions}
\begin{align}
	\label{Gamma_conditions_FDT}
		D_{ij}(\mathbf{q})=T\,\Gamma	_{ij}(\mathbf{q}),
		\\
		\label{Gamma_conditions_symmetry}
		\Gamma_{ij}(\mathbf{q})=\Gamma_{ji}(\mathbf{q}).
\end{align}	
\end{subequations}
The coefficient $\Gamma_{ij}(\mathbf{q})$, which is directly related to the irreversible diffusion, is termed \textit{dissipative} response coefficient. In contrast, $R_{ij}(\mathbf{q})$, which is unrelated to the irreversible processes, is termed \textit{reactive} response coefficient.

Using Eqs.~\eqref{Gamma_conditions_FDT} and \eqref{Gamma_conditions_symmetry}, the coefficient of each $\frac{\partial F}{\partial q_i} \frac{\partial F}{\partial q_j}$ in Eqs.~\eqref{FPE_expand} also vanishes. Next, if we define for convenience
\begin{align}
\label{psi_i_1_def}
	\psi_i^{\mathrm{(1)}}:=\psi_i-\sum_j\frac{\partial D_{ij}}{\partial q_j},
\end{align}
the remainder of Eq.~\eqref{FPE_expand} can be written as
\begin{align}
	\label{psi_i_1_procedure}
	\sum_i\left[-\frac{\partial \psi_i^\mathrm{(1)}}{\partial q_i}+\frac{1}{T}\left(\psi_i^\mathrm{(1)}-T\sum_j\frac{\partial R_{ij}}{\partial q_j}\right)\frac{\partial F}{\partial q_i}\right]=0,
\end{align}
where the antisymmetry $R_{ij} = -R_{ji}$ has been used. This implies
\begin{align}
	\label{psi_i_1_choice}
\psi_i^{\mathrm{(1)}}=T\sum_j\frac{\partial R_{ij}}{\partial q_j},
\end{align}
which also guarantees that the first term of Eq.~\eqref{psi_i_1_procedure} vanishes because
\begin{align}
	\label{psi_i_1_finalcheck}
	\sum_i\frac{\partial \psi_i^{\mathrm{(1)}}}{\partial q_i}=T\sum_{i,j}\frac{\partial R_{ij}}{\partial q_i \partial q_j}
	=\frac{T}{2}\sum_{i,j}\frac{\partial^2 (R_{ij}+R_{ji})}{\partial q_i \partial q_j}=0.
\end{align}

Combining Eqs.~\eqref{Gamma_conditions_FDT}, \eqref{psi_i_1_def}, and \eqref{psi_i_1_choice}, we identify
\begin{align}
	\label{psi_i_1_final}
	\psi_i=T\sum_j\frac{\partial \left(\Gamma_{ij}+R_{ij}\right)}{\partial q_j}.
\end{align}
This term is related to the {\em spurious drift}, alternatively called the {\em noise-induced drift} \cite{volpe2016effective}, which appears when the diffusivity is state dependent. Our result shows that the reactive response coefficients must also contribute to the spurious drift for equilibration.

Next, we require that the system must satisfy {\em detailed balance} (DB). Revisiting the FPE in Eq.~\eqref{eq:FPE}, and noting that the {\em mirror symmetry} $p_{\textrm{s}}(\mathcal{E}\mathbf{q})=p_{\textrm{s}}(\mathbf{q})$ is guaranteed for a diffusive system at equilibrium~\cite{yeo2016housekeeping}, the DB is equivalent to the following two conditions \cite{gardiner2009stochastic}:
\begin{subequations}\label{DB}
	\begin{align}
	&\epsilon_i \epsilon_j D_{ij} (\mathcal{E} \mathbf{q})
=D_{ij}(\mathbf{q}),\label{DB2}
\\ \nonumber
\\
	    &\epsilon_i A_i(\mathcal{E}\mathbf{q})p_{\textrm{s}}(\mathbf{q})
    =-A_i(\mathbf{q})p_{\textrm{s}}(\mathbf{q})
+2\sum_j\frac{\partial}{\partial q_j}\Big[D_{ij}(\mathbf{q})p_{\textrm{s}}(\mathbf{q})\Big].\label{DB1}
\end{align}
\end{subequations}
We note in passing that the mirror symmetry may not hold in the nonequlibrium steady state of the fuel-driven AOUP discussed in the main text.

From Eqs.~\eqref{Gamma_conditions_FDT} and \eqref{DB2}, the time-reversal property of the dissipative response coefficients is immediately determined as
\begin{align}
	\label{eq:gamma_ij_timerev}
	\Gamma_{ij}(\mathbf{q})=\epsilon_i \epsilon_j \Gamma_{ij}(\mathcal{E}\mathbf{q}).
\end{align}
Next, using Eqs.~\eqref{drift_linear}, \eqref{Gamma_conditions_FDT}, and \eqref{eq:gamma_ij_timerev} in Eq.~\eqref{DB1}, we also obtain the time-reversal property of the reactive response coefficients:
\begin{align}
	\label{eq:R_ij_timerev}
	R_{ij}(\mathbf{q})=-\epsilon_i \epsilon_j R_{ij}(\mathcal{E}\mathbf{q}).
\end{align}

Collecting Eqs.~\eqref{Gamma_conditions}, \eqref{psi_i_1_choice}, \eqref{eq:gamma_ij_timerev}, and \eqref{eq:R_ij_timerev}, we obtain the Langevin equation shown in Eq.~\eqref{general_recipe} and the Onsager reciprocal relations shown in Eq.~\eqref{eq:onsager} of the main text.

Some extra comments on why $\Gamma_{ij}$ ($R_{ij}$) can be called the dissipative (reactive) response coefficient are in order. Under time reversal, the flux $\dot{q}_i$ transforms as
\begin{align}
\dot{q}_i=\frac{dq_i}{dt}\quad\rightarrow\quad-\epsilon_i\dot{q}_i.
\end{align}
Meanwhile, noting the mirror symmetry $F(\mathcal{E}\mathbf{q})=F(\mathbf{q})$ and the time-reversal properties of the response coefficients shown in Eqs.~\eqref{eq:gamma_ij_timerev} and \eqref{eq:R_ij_timerev}, the two response components transform as
\begin{subequations}
	\begin{align}
	-R_{ij}(\mathbf{q})\frac{\partial F(\mathbf{q})}{\partial q_j}\quad\rightarrow\quad\epsilon_i R_{ij}(\mathbf{q})\frac{\partial F(\mathbf{q})}{\partial q_j},
\\ \nonumber
\\
	-\Gamma_{ij}(\mathbf{q})\frac{\partial F(\mathbf{q})}{\partial q_j}\quad\rightarrow\quad-\epsilon_i\Gamma_{ij}(\mathbf{q})\frac{\partial F(\mathbf{q})}{\partial q_j}.
\end{align}
\end{subequations}
This shows that the reactive response coefficient $R_{ij}$ (dissipative response coefficient $\Gamma_{ij}$) is associated with a response term whose sign under time reversal is the same as (opposite to) that of the flux $\dot{q}_i$. Thus, the reactive (dissipative) response coefficient accounts for the reversible (irreversible) part of the dynamics.

\section{Derivation of the steady-state solution}

\subsection{Derivation of the steady-state distribution}
\label{apdx_covmtx}

Here we present an explicit derivation of the steady-state solution for the active heat engine described by Eq.~\eqref{eq:gyrator}. Since the dynamics is fully linear, the statistics of $\mathbf{r}$ is given by a multivariate Gaussian distribution of zero mean. Thus, by obtaining the steady-state covariance matrix $\mathbb{C}:=\langle\mathbf{r}\,\mathbf{r}^{\textrm{T}}\rangle$, we fully characterize the steady-state distribution. Thus, in the following, we show how to derive $\mathbb{C}$. Note that in this subsection and the next we stick to the notations for the even-parity engine. As stated in the main text, the results for the odd-parity engine can be obtained by a simple change of variables.

First, using Eq.~\eqref{eq:gyrator}, we obtain
\begin{align}
	\langle \mathbf{r} \circ \dot{\mathbf{r}}^\mathrm{T}\rangle=-\langle \mathbf{r} \,\mathbf{r}^\mathrm{T}\mathbb{K}^\mathrm{T}\rangle\,+\,\langle\mathbf{r}\circ\bm{\xi}^\mathrm{T}\rangle
	\\ \nonumber
	=-\mathbb{C}\mathbb{K}^\mathrm{T}+\langle\mathbf{r}\circ\bm{\xi}^\mathrm{T}\rangle.
\end{align}
Using the formal solution of Eq.~\eqref{eq:gyrator}, one can show the relation $\langle \mathbf{r}\circ\bm{\xi}^\mathrm{T} \rangle = \mathbb{D}$. This leads to
\begin{align}
	\label{covmtxA}
	\langle {\mathbf{r}} \circ \dot{{\mathbf{r}}}^\textrm{T}\rangle=-\mathbb{C}\mathbb{K}^\textrm{T}+\mathbb{D}.
\end{align}
In the steady state,
\begin{align}
	\frac{d}{dt} \langle \mathbf{r} \mathbf{r}^\mathrm{T} \rangle = \langle {\mathbf{r}} \circ \dot{{\mathbf{r}}}^\textrm{T}\rangle + \langle {\mathbf{r}} \circ \dot{{\mathbf{r}}}^\textrm{T}\rangle^\mathrm{T} = 0.
\end{align}
Using Eq.~\eqref{covmtxA} in this relation, we obtain
\begin{align}
	\label{covmtxB}
    \mathbb{K}\mathbb{C}+\mathbb{C}\mathbb{K}^\textrm{T}=2\mathbb{D},
\end{align}
which is known as the {\em Lyapunov equation}.

In our case, this equation can be solved more easily using the hierarchical structure of $\mathbb{K}$. First, since the equations for the third and fourth components ($x_1$ and $x_2$) are isolated from the others, we immediately obtain
\begin{align}
	C_{33}=\frac{\tau_1T_1}{\gamma_1},\quad
C_{34}=0,\quad
C_{44}=\frac{\tau_2T_2}{\gamma_2}.
\label{ss_sols_obvious}
\end{align}
Using these, the other components of Eq.~\eqref{covmtxB} can be written as follows:
\begin{subequations}\label{covmtx_eqn}
\begin{align}
-K C_{11}+\lambda_1 C_{12} + \zeta_1\Delta\mu_1 C_{13}
+ T_1=0,
\label{covmtx_eqn1}
\\
\nonumber\\
-2KC_{12}+\lambda_1 C_{22} + \zeta_1\Delta\mu_1 C_{23}\qquad\qquad
\nonumber\\
+\lambda_2 C_{11} + \zeta_2\Delta\mu_2 C_{14}=0,
\label{covmtx_eqn2}
\\
\nonumber\\
\Big(\frac{\Gamma}{\tau_1}+K\Big)C_{13}=\lambda_1 C_{23} +\zeta_1\Delta\mu_1\frac{\tau_1T_1}{\gamma_1},
\label{covmtx_eqn3}
\\
\nonumber\\
\Big(\frac{\Gamma}{\tau_2}+K\Big)C_{14}=\lambda_1 C_{24},
\label{covmtx_eqn4}
\\
\nonumber\\
\lambda_2 C_{12} - K C_{22} + \zeta_2\Delta\mu_2 C_{24}+T_2=0,
\label{covmtx_eqn5}
\\
\nonumber\\
\Big(\frac{\Gamma}{\tau_1}+K\Big)C_{23}=\lambda_2 C_{13},
\label{covmtx_eqn6}
\\
\nonumber\\
\Big(\frac{\Gamma}{\tau_2}+K\Big)C_{24}=\lambda_2 C_{14} +\zeta_2\Delta\mu_2\frac{\tau_2T_2}{\gamma_2}.
\label{covmtx_eqn7}
\end{align}
\end{subequations}
Putting Eqs.~\eqref{covmtx_eqn3} and \eqref{covmtx_eqn6} together, we get
\begin{subequations}
\label{C13_C23}
\begin{align}
	C_{13}
=
\zeta_1\Delta\mu_1 \frac{\tau_1T_1}{\gamma_1}
\frac{
(\Gamma / \tau_1)+K
}{
\big[(\Gamma / \tau_1)+K\big]^2
-\lambda_1\lambda_2
},
    \label{C13}
\\
C_{23}=
\zeta_1\Delta\mu_1 \frac{\tau_1T_1}{\gamma_1}
\frac{\lambda_2}{
\big[(\Gamma / \tau_1)+K\big]^2
-\lambda_1\lambda_2
}.
    \label{C23}
\end{align}
\end{subequations}
Next, similarly from Eqs.~\eqref{covmtx_eqn4} and \eqref{covmtx_eqn7}, we obtain
\begin{subequations}
\label{C24_C14}
\begin{align}
	    C_{24}
=
\zeta_2\Delta\mu_2 \frac{\tau_2T_2}{\gamma_2}
\frac{
(\Gamma / \tau_2)+K
}{
\big[(\Gamma / \tau_2)+K\big]^2
-\lambda_1\lambda_2
},
    \label{C24}
\\
    C_{14}=
\zeta_2\Delta\mu_2 \frac{\tau_2T_2}{\gamma_2}
\frac{\lambda_1}{
\big[(\Gamma / \tau_2)+K\big]^2
-\lambda_1\lambda_2
}.
    \label{C14}
\end{align}
\end{subequations}

Using these relations in the remaining Eqs.~\eqref{covmtx_eqn1}, \eqref{covmtx_eqn2}, and \eqref{covmtx_eqn5}, we obtain
\begin{widetext}
\begin{align}
	C_{12}=
\frac{1}{K^2-\lambda_1\lambda_2}
\bigg[
\frac{\lambda_2 T_1 + \lambda_1 T_2}{2}
+
\frac{\tau_1T_1}{\gamma_1} \frac{(\zeta_1\Delta\mu_1)^2~(K+\frac{\Gamma}{2\tau_1})\lambda_2}{(\frac{\Gamma}{\tau_1}+K)^2-\lambda_1\lambda_2}
+
\frac{\tau_2T_2}{\gamma_2} \frac{(\zeta_2\Delta\mu_2)^2~(K+\frac{\Gamma}{2\tau_2})\lambda_1}{(\frac{\Gamma}{\tau_2}+K)^2-\lambda_1\lambda_2}
\bigg]
    \label{C12}
\end{align}
as well as
\begin{subequations}
    \label{C11_C22}
\begin{align}
	C_{11}&=\frac{1}{K}\big[
\lambda_1 C_{12} + (\zeta_1\Delta\mu_1) C_{13} + T_1
\big],
    \label{C11}
\\
C_{22}&=\frac{1}{K}\big[
\lambda_2 C_{12} + (\zeta_2\Delta\mu_2) C_{24} + T_2
\big].
    \label{C22}
\end{align}
\end{subequations}
\end{widetext}

\onecolumngrid
\subsection{Derivation of the mean energy currents}
\label{apdx_explform}

Using the steady-state statistics obtained above, here we explicitly derive the mean energy currents $\langle \dot{W}_\mathrm{out} \rangle$ and $\langle \dot{W}_{\mathrm{chem},i} \rangle$. First, taking advantage of the steady-state condition
\begin{align}
\frac{d}{dt}\,\langle X_1 X_2\rangle
=\langle X_1 %\circ
\dot{X}_2\rangle+\langle X_2 %\circ
\dot{X}_1\rangle=0,
\end{align}
the mean extracted power $\langle \dot{W}_\mathrm{out} \rangle = \langle \dot{W}_\mathrm{out,1}\rangle+\langle \dot{W}_\mathrm{out,2}\rangle$ given by Eq.~\eqref{eq:wout_stoc} of the main text can be written as $\langle\dot{W}_{\textrm{out}}\rangle=(\lambda_1-\lambda_2)\langle X_1 %\circ 
\dot{X}_2\rangle$ in the steady state. From this, using Eq.~\eqref{covmtxA} and the steady-state statistics obtained above, we obtain the explicit formula
\begin{align}
	    \label{wout}
    \langle\dot{W}_{\textrm{out}}\rangle=\frac{\lambda_1-\lambda_2}{2K}\bigg[
    \frac{\lambda_2T_1-\lambda_1T_2}{\Gamma}+\frac{T_1 \tau_1^2\,\lambda_1}{(\Gamma+K\tau_1)^2-\tau_1^2\lambda_1\lambda_2}(\zeta_1\Delta\mu_1)^2-\frac{T_2 \tau_2^2 \,\lambda_2}{(\Gamma+K\tau_2)^2-\tau_2^2\lambda_1\lambda_2}(\zeta_2\Delta\mu_2)^2
    \bigg].
\end{align}
is obtained. When $\lambda_1 > \lambda_2$, the chemical driving $\Delta\mu_1$ associated with $X_1$ positively contributes to the power, while $\Delta\mu_2$ deteriorates the power. The reparametrization introduced in the Sec.~\ref{sec:emp} of the main text gives us an alternative expression
\begin{align}
	\label{wout_rc}
    \begin{split}
\langle\dot{W}_{\textrm{out}}\rangle&=
\,c\Big(1-\frac{1}{r}\Big)g_1(c)-\,c\big(r-1\big)g_2(c)+c\big(r-1\big)\Big(\frac{1}{r}-\frac{T_2}{T_1}\Big)\frac{T_1}{2\Gamma K}\quad\quad\quad\quad\quad\quad\quad\quad\quad
\\
&=c\bigg\{
g_1(c)+g_2(c)+\frac{T_1+T_2}{2\Gamma K}-\Big[
\Big(g_2(c) + \frac{T_2}{2\Gamma K}\Big)r
+
\Big(g_1(c) + \frac{T_1}{2\Gamma K}\Big)\frac{1}{r}
\Big]
\bigg\},
    \end{split}
\end{align}
where
\begin{align}
	g_i(c)
\equiv\frac{\tau_i^2}{2K}\,\frac{T_i}{\gamma_i}\,\frac{(\zeta_i\Delta\mu_i)^2}{(\Gamma+K\tau_i)^2-\tau_i^2c}.
\end{align}
To maximize the power by varying $r$ for a given value of $c$, we can use the inequality between the arithmetic and the geometric means to the second line of Eq.~\eqref{wout_rc}, which yields the maximum power
\begin{align}
	P^*\equiv\left.\langle\dot{W}_{\textrm{out}}\rangle\right|_{r=r^*}=c\left(\sqrt{g_1(c)+\frac{T_1}{2\Gamma K}}-\sqrt{g_2(c)+\frac{T_2}{2\Gamma K}}\right)^2
\end{align}
along with the optimal value of $r$ given by
\begin{align}
r^* = \sqrt{\Big(g_1(c) + \frac{T_1}{2\Gamma K}\Big)/\Big(g_2(c) + \frac{T_2}{2\Gamma K}\Big)}.
\end{align}
In the main text, the values of these quantities are given for the two special cases $\Delta\mu_1=0$ and $\Delta\mu_2=0$.

Meanwhile, the average chemical work rate associated with $X_i$ can be calculated explicitly by applying the exact statistics obtained above to Eqs.~\eqref{eq:even_Wchem} and \eqref{eq:odd_Wchem}. For the even-parity case, the quantity reads:
\begin{align}
\label{w_chem_avg_even}
	    \langle\dot{W}_{\textrm{chem},i}\rangle
        =
        \frac{T_i}{\gamma_i}\frac{
        (\zeta_i\Delta\mu_i)^2
        }{
        (\Gamma+K\tau_i)^2-\tau_i^2\,c
        }\,
        (\Gamma + K\tau_i)\tau_i.
\end{align}
The corresponding quantity for the odd-parity case is 
\begin{align}
\label{w_chem_avg_odd}
        \langle\dot{W}_{\textrm{chem},i}\rangle
        =
        \gamma_i'T_i~ \frac{
        (\zeta_i'\Delta\mu_i)^2
        }{
        (\Gamma+K\tau_i)^2-\tau_i^2\,c
        }\,
        \frac{[\Gamma K + (K^2-c)\tau_i]\,\tau_i^2}{\Gamma}.
\end{align}
Note that these rates do not depend on $r$---the chemical work of our model is unrelated to the optimality of the engine. Also, they are always positive as long as the stability condition $c<K^2$ is valid. Plugging $P^*$ of the main text and Eqs.~\eqref{w_chem_avg_even} and \eqref{w_chem_avg_odd} to Eq.~\eqref{eq:eta_comp} gives the explicit formulae Eqs.~\eqref{eq:cEMP1} and \eqref{eq:cEMP2} for the EMP.

\section{The detailed structure of entropy production}
\label{apdx_bDB_as}
Here we provide detailed derivations of the EP components discussed in Sec.~\ref{sec:tighterbound} of the main text, following the decomposition scheme of \cite{yeo2016housekeeping}.
First, using the ``nonequilibrium potential'' $\phi(\mathbf{r})\equiv-\ln p_{\textrm{s}}(\mathbf{r})$ and its time-reversed counterpart $\phi^\mathrm{R}(\mathbf{r})\equiv-\ln p_s(\mathcal{E}\mathbf{r})$, we rewrite the steady-state condition $\nabla\cdot\mathbf{J}^{\textrm{s}}(\mathbf{r})=0$ for the probability current $J_i^\mathrm{s}(\mathbf{r})=(-\sum_j K_{ij}\,r_j)p_s(\mathbf{r})-\sum_j\partial_j\left(D_{ij}(\mathbf{r})~p_s(\mathbf{r})\right)$. For the even-parity case, the condition is given by
\begin{align}
	\label{eq:log_density_ss_even}
	0=\frac{1}{\tau_1}+\left(-\frac{1}{\tau_1}x_1\right)\frac{\partial\phi}{\partial x_1}+\frac{T_1}{\gamma_1}\left[-\frac{\partial^2\phi}{\partial x_1^2}+\left(\frac{\partial\phi}{\partial x_1}\right)^2\right]
	+ \frac{1}{\tau_2}+\left(-\frac{1}{\tau_2}x_2\right)\frac{\partial\phi}{\partial x_2}+\frac{T_2}{\gamma_2}\left[-\frac{\partial^2\phi}{\partial x_2^2}+\left(\frac{\partial\phi}{\partial x_2}\right)^2\right]
	 \nonumber
	\\
	+ \frac{K}{\Gamma}+\left(-\frac{K}{\Gamma}X_1+\frac{\lambda_1}{\Gamma}X_2+\frac{\zeta_1\Delta\mu_1}{\Gamma}x_1\right)\frac{\partial\phi}{\partial X_1}+\frac{T_1}{\Gamma}\left[-\frac{\partial^2\phi}{\partial X_1^2}+\left(\frac{\partial\phi}{\partial X_1}\right)^2\right]
	  \nonumber
	\\
	+ \frac{K}{\Gamma}+\left(-\frac{K}{\Gamma}X_2+\frac{\lambda_2}{\Gamma}X_1+\frac{\zeta_2\Delta\mu_2}{\Gamma}x_2\right)\frac{\partial\phi}{\partial X_2}+\frac{T_2}{\Gamma}\left[-\frac{\partial^2\phi}{\partial X_2^2}+\left(\frac{\partial\phi}{\partial X_2}\right)^2\right],
\end{align}
with the time-reversed counterpart
\begin{align}
	\label{eq:log_density_ss_even_R}
	0=\frac{1}{\tau_1}+\left(-\frac{1}{\tau_1}x_1\right)\frac{\partial\phi^\mathrm{R}}{\partial x_1}+\frac{T_1}{\gamma_1}\left[-\frac{\partial^2\phi^\mathrm{R}}{\partial x_1^2}+\left(\frac{\partial\phi^\mathrm{R}}{\partial x_1}\right)^2\right]
	+ \frac{1}{\tau_2}+\left(-\frac{1}{\tau_2}x_2\right)\frac{\partial\phi^\mathrm{R}}{\partial x_2}+\frac{T_2}{\gamma_2}\left[-\frac{\partial^2\phi^\mathrm{R}}{\partial x_2^2}+\left(\frac{\partial\phi^\mathrm{R}}{\partial x_2}\right)^2\right]
	 \nonumber
	\\
	+ \frac{K}{\Gamma}+\left(-\frac{K}{\Gamma}X_1+\frac{\lambda_1}{\Gamma}X_2+\frac{\zeta_1\Delta\mu_1}{\Gamma}x_1\right)\frac{\partial\phi^\mathrm{R}}{\partial X_1}+\frac{T_1}{\Gamma}\left[-\frac{\partial^2\phi^\mathrm{R}}{\partial X_1^2}+\left(\frac{\partial\phi^\mathrm{R}}{\partial X_1}\right)^2\right]
	  \nonumber
	\\
	+ \frac{K}{\Gamma}+\left(-\frac{K}{\Gamma}X_2+\frac{\lambda_2}{\Gamma}X_1+\frac{\zeta_2\Delta\mu_2}{\Gamma}x_2\right)\frac{\partial\phi^\mathrm{R}}{\partial X_2}+\frac{T_2}{\Gamma}\left[-\frac{\partial^2\phi^\mathrm{R}}{\partial X_2^2}+\left(\frac{\partial\phi^\mathrm{R}}{\partial X_2}\right)^2\right].
\end{align}
For the odd-parity case, we similarly obtain
\begin{align}
	\label{eq:log_density_ss_odd}
	0=\frac{1}{\tau_1}+\left(-\frac{1}{\tau_1}p_1\right)\frac{\partial\phi}{\partial p_1}+\gamma_1'T_1\left[-\frac{\partial^2\phi}{\partial p_1^2}+\left(\frac{\partial\phi}{\partial p_1}\right)^2\right]
	+ \frac{1}{\tau_2}+\left(-\frac{1}{\tau_2}p_2\right)\frac{\partial\phi}{\partial p_2}+\gamma_2'T_2\left[-\frac{\partial^2\phi}{\partial p_2^2}+\left(\frac{\partial\phi}{\partial p_2}\right)^2\right]
	 \nonumber
	\\
	+ \frac{K}{\Gamma}+\left(-\frac{K}{\Gamma}X_1+\frac{\lambda_1}{\Gamma}X_2+\frac{\zeta_1'\Delta\mu_1}{\Gamma}p_1\right)\frac{\partial\phi}{\partial X_1}+\frac{T_1}{\Gamma}\left[-\frac{\partial^2\phi}{\partial X_1^2}+\left(\frac{\partial\phi}{\partial X_1}\right)^2\right]
	  \nonumber
	\\
	+ \frac{K}{\Gamma}+\left(-\frac{K}{\Gamma}X_2+\frac{\lambda_2}{\Gamma}X_1+\frac{\zeta'_2\Delta\mu_2}{\Gamma}p_2\right)\frac{\partial\phi}{\partial X_2}+\frac{T_2}{\Gamma}\left[-\frac{\partial^2\phi}{\partial X_2^2}+\left(\frac{\partial\phi}{\partial X_2}\right)^2\right],
\end{align}
\begin{align}
	\label{eq:log_density_ss_odd_R}
	0=\frac{1}{\tau_1}+\left(-\frac{1}{\tau_1}p_1\right)\frac{\partial\phi^\mathrm{R}}{\partial p_1}+\gamma_1'T_1\left[-\frac{\partial^2\phi^\mathrm{R}}{\partial p_1^2}+\left(\frac{\partial\phi^\mathrm{R}}{\partial p_1}\right)^2\right]
	+ \frac{1}{\tau_2}+\left(-\frac{1}{\tau_2}p_2\right)\frac{\partial\phi^\mathrm{R}}{\partial p_2}+\gamma_2'T_2\left[-\frac{\partial^2\phi^\mathrm{R}}{\partial p_2^2}+\left(\frac{\partial\phi^\mathrm{R}}{\partial p_2}\right)^2\right]
	 \nonumber
	\\
	+ \frac{K}{\Gamma}+\left(-\frac{K}{\Gamma}X_1+\frac{\lambda_1}{\Gamma}X_2-\frac{\zeta_1'\Delta\mu_1}{\Gamma}p_1\right)\frac{\partial\phi^\mathrm{R}}{\partial X_1}+\frac{T_1}{\Gamma}\left[-\frac{\partial^2\phi^\mathrm{R}}{\partial X_1^2}+\left(\frac{\partial\phi^\mathrm{R}}{\partial X_1}\right)^2\right]
	  \nonumber
	\\
	+ \frac{K}{\Gamma}+\left(-\frac{K}{\Gamma}X_2+\frac{\lambda_2}{\Gamma}X_1-\frac{\zeta_2'\Delta\mu_2}{\Gamma}p_2\right)\frac{\partial\phi^\mathrm{R}}{\partial X_2}+\frac{T_2}{\Gamma}\left[-\frac{\partial^2\phi^\mathrm{R}}{\partial X_2^2}+\left(\frac{\partial\phi^\mathrm{R}}{\partial X_2}\right)^2\right].
\end{align}

Now we introduce a measure of the mirror symmetry breaking $\phi^{\textrm{A}}(\mathbf{r})\equiv\phi(\mathbf{r})-\phi(\mathcal{E}\mathbf{r})$ and a mixture distribution
\begin{align}
	\label{eq:psi_sigma}
	\psi_\sigma(\mathbf{r})\equiv\sigma\phi(\mathbf{r})+(1-\sigma)\phi(\mathcal{E}\mathbf{r})
	=\phi(\mathbf{r})-(1-\sigma)\phi^{\textrm{A}}(\mathbf{r})
	=\phi(\mathcal{E}\mathbf{r})+\sigma\phi^{\textrm{A}}(\mathbf{r}),
\end{align}
where $\sigma$ is a real-valued parameter characterizing the generalized adjoint process associated with the housekeeping EP~\cite{yeo2016housekeeping}.

\subsection{Even-parity case}
For the even-parity case, the mirror symmetry $\phi^\mathrm{A}(\mathbf{r}) = 0$ trivially holds
%, {\em i.e.}, $\phi(\mathcal{E}\mathbf{r})=\phi(\mathbf{r})$
since there are no odd-parity variables. Thus $\psi_{\sigma}(\mathbf{r})=\phi(\mathbf{r})$, resulting in the following (also see Eq.~(71) of \cite{yeo2016housekeeping}):
\begin{align}
	\label{eq:S_bDB_even_first}
\dot{S}_{\textrm{bDB}}%^{\textrm{(e)}}
&=
-\frac{1}{T_1}\dot{Q}_{1}%^{\textrm{(e)}}
-\frac{1}{T_2}\dot{Q}_{2}%^{\textrm{(e)}}
+\bigg\{\frac{\partial \phi}{\partial X_1}\circ \dot{X}_1
+\frac{\partial \phi}{\partial X_2}\circ \dot{X}_2
+\frac{\partial \phi}{\partial x_1}\circ \dot{x}_1
+\frac{\partial \phi}{\partial x_2}\circ \dot{x}_1
\bigg\}
\nonumber \\
&+\bigg\{
\Big[
\Big(-\frac{1}{\tau_1}x_1\Big)\frac{\partial \phi}{\partial x_1}
+\frac{T_1}{\gamma_1}\Big(\frac{\partial \phi}{\partial x_1}\Big)^2
+\frac{1}{\tau_1}
-\frac{T_1}{\gamma_1}\frac{\partial^2 \phi}{\partial x_1^2}
\Big]
+
\Big[
\Big(-\frac{1}{\tau_2}x_2\Big)\frac{\partial \phi}{\partial x_2}
+\frac{T_2}{\gamma_2}\Big(\frac{\partial \phi}{\partial x_2}\Big)^2
+\frac{1}{\tau_2}
-\frac{T_2}{\gamma_2}\frac{\partial^2 \phi}{\partial x_2^2}
\Big]
\nonumber \\
&
\qquad\qquad\qquad\qquad\qquad\qquad
+
\Big[
\Big(-\frac{K}{\Gamma}X_1
+\frac{\lambda_1}{\Gamma}X_2+\frac{\zeta_1\Delta\mu_1}{\Gamma}x_1\Big)\frac{\partial \phi}{\partial X_1}+\frac{T_1}{\Gamma}\Big(\frac{\partial\phi}{\partial X_1}\Big)^2
+\frac{K}{\Gamma}-\frac{T_1}{\Gamma}\frac{\partial^2 \phi}{\partial X_1^2}
\Big]
\nonumber \\
&
\qquad\qquad\qquad\qquad\qquad\qquad
+
\Big[
\Big(-\frac{K}{\Gamma}X_2
+\frac{\lambda_2}{\Gamma}X_1
+\frac{\zeta_2\Delta\mu_2}{\Gamma}x_2\Big)\frac{\partial \phi}{\partial X_2}+\frac{T_2}{\Gamma}\Big(\frac{\partial\phi}{\partial X_2}\Big)^2
+\frac{K}{\Gamma}-\frac{T_2}{\Gamma}\frac{\partial^2 \phi}{\partial X_2^2}
\Big]
\bigg\}.
\end{align}
Comparing this with Eq.~\eqref{eq:log_density_ss_even}, we see that the expression inside the second curly bracket is zero. In addition, the first curly bracket is readily identified as the total time derivative of $\phi$. Thus we obtain 
\begin{align}
	\label{S_bDB_even_simple}
	\dot{S}_{\mathrm{bDB}}=-\frac{1}{T_1}\dot{Q}_1-\frac{1}{T_2}\dot{Q}_2+\dot{\phi}=\dot{S}_{\mathrm{env}}+\dot{\phi}=\dot{S}_{\mathrm{tot}}.
\end{align}
The last equality comes from the identification of $\phi = -\ln p_s$ as the stochastic entropy of the system. Meanwhile, the EP rate associated with the breaking of the mirror symmetry is trivially given by $\dot{S}_{\textrm{as}}=0$ for the even-parity case since $\phi^\mathrm{A}(\mathbf{r})=0$ (see Eq.~(78) of \cite{yeo2016housekeeping}). This result is also in agreement with Eq.~\eqref{S_bDB_even_simple}, which means that the breaking of DB accounts for the EP entirely.

\subsection{Odd-parity case}
The housekeeping EP has more detailed structure in the odd-parity case. Again referring to Eq.~(71) of \cite{yeo2016housekeeping},
\begin{align}
	\label{eq:S_bDB_odd_first}
	&\dot{S}_{\textrm{bDB}}%^{\textrm{(o)}}
    =
    -\frac{1}{T_1}\dot{Q}_1%^{\textrm{(o)}}
    -\frac{1}{T_2}\dot{Q}_2%^{\textrm{(o)}}
    +
    \bigg\{
    \frac{\partial\psi_{\sigma}}{\partial X_1}\circ\dot{X}_1
    +\frac{\partial\psi_{\sigma}}{\partial X_2}\circ\dot{X}_2
    +\frac{\partial\psi_{\sigma}}{\partial p_1}\circ\dot{p}_1
    +\frac{\partial\psi_{\sigma}}{\partial p_2}\circ\dot{p}_2
    \bigg\}
        \nonumber
    \\
    &
    +
    \bigg\{
    \Big[
    \Big(-\frac{p_1}{\tau_1}\Big)\frac{\partial \psi_{\sigma}}{\partial p_1}
    +\gamma_1'T_1\Big(\frac{\partial \psi_{\sigma}}{\partial p_1}\Big)^2
    +\frac{1}{\tau_1}-\gamma_1'T_1\frac{\partial^2 \psi_{\sigma}}{\partial p_1^2}
    \Big]
    +
    \Big[
    \Big(-\frac{p_2}{\tau_2}\Big)\frac{\partial \psi_{\sigma}}{\partial p_2}
    +\gamma_2'T_2\Big(\frac{\partial \psi_{\sigma}}{\partial p_2}\Big)^2+\frac{1}{\tau_2}-\gamma_2'T_2\frac{\partial^2 \psi_{\sigma}}{\partial p_2^2}
    \Big]
            \nonumber
    \\
    &
    \qquad\qquad\qquad\qquad\qquad\qquad\qquad\quad
    +
    \Big[
    \Big(-\frac{K}{\Gamma}X_1+\frac{\lambda_1}{\Gamma}X_2-\frac{\zeta'_1\Delta\mu_1}{\Gamma}p_1\Big)\frac{\partial \psi_{\sigma}}{\partial X_1}
    +\frac{T_1}{\Gamma}\Big(\frac{\partial\psi_{\sigma}}{\partial X_1}\Big)^2+\frac{K}{\Gamma}-\frac{T_1}{\Gamma}\frac{\partial^2\psi_{\sigma}}{\partial X_1^2}
    \Big]
    \nonumber
    \\
    &
    \qquad\qquad\qquad\qquad\qquad\qquad\qquad\quad
    +
    \Big[
    \Big(-\frac{K}{\Gamma}X_2+\frac{\lambda_2}{\Gamma}X_1-\frac{\zeta'_2\Delta\mu_2}{\Gamma}p_2\Big)\frac{\partial \psi_{\sigma}}{\partial X_2}
    +\frac{T_2}{\Gamma}\Big(\frac{\partial\psi_{\sigma}}{\partial X_2}\Big)^2+\frac{K}{\Gamma}-\frac{T_2}{\Gamma}\frac{\partial^2\psi_{\sigma}}{\partial X_2^2}
    \Big]
    \bigg\}.
\end{align}
Using the relation $\psi_{\sigma}(\mathbf{r})=\phi^{\textrm{R}}(\mathbf{r})+\sigma\phi^{\textrm{A}}(\mathbf{r})$, this can be rewritten as
\begin{align}
	\label{eq:S_bDB_odd_simple}
	\dot{S}_\mathrm{bDB}=\left(-\frac{1}{T_1}\dot{Q}_1-\frac{1}{T_2}\dot{Q}_2\right)+\dot{\psi}^\sigma+\sigma^2 A+\sigma B
	%\nonumber \\&
	\,\,&=\,\,\dot{S}_{\mathrm{env}}+\dot{\phi}-(1-\sigma)\dot{\phi}^{\mathrm{A}}+\sigma^2 A+\sigma B
	\nonumber \\
	&=\dot{S}_\mathrm{tot}-(1-\sigma)\dot{\phi}^\mathrm{A}+\sigma^2 A + \sigma B,
\end{align}
where
\begin{equation}
    \label{A}
A\equiv\frac{T_1}{\Gamma}\big(\frac{\partial\phi^{\textrm{A}}}{\partial X_1}\big)^2
+\frac{T_2}{\Gamma}\big(\frac{\partial\phi^{\textrm{A}}}{\partial X_2}\big)^2
+\gamma_1'T_1\big(\frac{\partial\phi^{\textrm{A}}}{\partial p_1}\big)^2
+\gamma_2'T_2\big(\frac{\partial\phi^{\textrm{A}}}{\partial p_2}\big)^2,
\end{equation}
\begin{equation}
\begin{split}
    \label{B}
B\equiv\Big[-\frac{1}{\tau_1}p_1 + 2\gamma_1'T_1 \big(\frac{\partial \phi^{\textrm{R}}}{\partial p_1}\big)\Big]\frac{\partial \phi^{\textrm{A}}}{\partial p_1}
-\gamma_1'T_1\frac{\partial^2 \phi^{\textrm{A}}}{\partial p_1^2}
+
\Big[-\frac{1}{\tau_2}p_2 + 2\gamma_2'T_2 \big(\frac{\partial \phi^{\textrm{R}}}{\partial p_2}\big)\Big]\frac{\partial \phi^{\textrm{A}}}{\partial p_2}
-\gamma_2'T_2\frac{\partial^2 \phi^{\textrm{A}}}{\partial p_2^2}\\
+
\Big[
-\frac{K}{\Gamma}X_1+\frac{\lambda_1}{\Gamma}X_2-\frac{\zeta'_1\Delta\mu_1}{\Gamma}p_1 + \frac{2T_1}{\Gamma}\big(\frac{\partial \phi^{\textrm{R}}}{\partial X_1}\big)
\Big]\frac{\partial \phi^{\textrm{A}}}{\partial X_1} - \frac{T_1}{\Gamma}\frac{\partial^2 \phi^{\textrm{A}}}{\partial X_1^2}\\
+
\Big[
-\frac{K}{\Gamma}X_2+\frac{\lambda_2}{\Gamma}X_1-\frac{\zeta'_2\Delta\mu_2}{\Gamma}p_2
+\frac{2T_2}{\Gamma}\big(\frac{\partial \phi^{\textrm{R}}}{\partial X_2}\big)
\Big]\frac{\partial \phi^{\textrm{A}}}{\partial X_2} - \frac{T_2}{\Gamma}\frac{\partial^2 \phi^{\textrm{A}}}{\partial X_2^2}.
\end{split}
\end{equation}

To simplify this further, we revisit the steady-state conditions. Subtracting Eq.~\eqref{eq:log_density_ss_odd_R} from Eq.~\eqref{eq:log_density_ss_odd} side by side, we get
\begin{align}
0&=\left(-\frac{p_1}{\tau_1}\right)\frac{\partial\phi^\mathrm{A}}{\partial p_1}+\gamma_1'T_1\left[-\frac{\partial^2\phi^\mathrm{A}}{\partial p_1^2}
+\left(\frac{\partial\phi}{\partial p_1}\right)^2-\left(\frac{\partial\phi^\mathrm{R}}{\partial p_1}\right)^2
\right]
+
\left(-\frac{p_2}{\tau_2}\right)\frac{\partial\phi^\mathrm{A}}{\partial p_2}+\gamma_2'T_2\left[-\frac{\partial^2\phi^\mathrm{A}}{\partial p_2^2}
+\left(\frac{\partial\phi}{\partial p_2}\right)^2-\left(\frac{\partial\phi^\mathrm{R}}{\partial p_2}\right)^2
\right]
\nonumber \\
&
\qquad\qquad\qquad\quad+
\left(-\frac{K}{\Gamma}X_1+\frac{\lambda_1}{\Gamma}X_2-\frac{\zeta_1'\Delta\mu_1}{\Gamma}p_1\right)\frac{\partial\phi^\mathrm{A}}{\partial X_1}+2\frac{\zeta_1'\Delta\mu_1}{\Gamma}p_1\frac{\partial\phi}{\partial X_1}+\frac{T_1}{\Gamma}\left[-\frac{\partial^2\phi^\mathrm{A}}{\partial X_1^2}+\left(\frac{\partial\phi}{\partial X_1}\right)^2-\left(\frac{\partial\phi^\mathrm{R}}{\partial X_1}\right)^2\right]
\nonumber \\
&
\qquad\qquad\qquad\quad+
\left(-\frac{K}{\Gamma}X_2+\frac{\lambda_2}{\Gamma}X_1-\frac{\zeta_2'\Delta\mu_2}{\Gamma}p_2\right)\frac{\partial\phi^\mathrm{A}}{\partial X_2}+2\frac{\zeta_2'\Delta\mu_2}{\Gamma}p_2\frac{\partial\phi}{\partial X_2}+\frac{T_2}{\Gamma}\left[-\frac{\partial^2\phi^\mathrm{A}}{\partial X_2^2}+\left(\frac{\partial\phi}{\partial X_2}\right)^2-\left(\frac{\partial\phi^\mathrm{R}}{\partial X_2}\right)^2\right]
\nonumber \\
&
=A+B+2\frac{\zeta_1'\Delta\mu_1}{\Gamma}p_1\frac{\partial\phi}{\partial X_1}
	+2\frac{\zeta_2'\Delta\mu_2}{\Gamma}
	p_2\frac{\partial\phi}{\partial X_2}.
	\label{eq:log_density_odd_diff}
\end{align}
Using this relation to eliminate $B$  in Eq.~\eqref{eq:S_bDB_odd_simple}, we obtain
\begin{align}
	\label{eq:S_bDB_odd_final}
	\dot{S}_{\mathrm{bDB}}&=\dot{S}_{\mathrm{tot}}-(1-\sigma)\dot{\phi}^\mathrm{A}-\sigma(1-\sigma)A -2\sigma\left(
	\frac{\zeta_1'\Delta\mu_1}{\Gamma}p_1\frac{\partial\phi}{\partial X_1}
	+\frac{\zeta_2'\Delta\mu_2}{\Gamma}
	p_2\frac{\partial\phi}{\partial X_2}
	\right).
\end{align}
This immediately implies
\begin{align}
	\label{eq:S_as_odd_final}
\dot{S}_\mathrm{as}=(1-\sigma)\dot{\phi}^\mathrm{A}+\sigma(1-\sigma)A+2\sigma\left(
	\frac{\zeta_1'\Delta\mu_1}{\Gamma}p_1\frac{\partial\phi}{\partial X_1}
	+\frac{\zeta_2'\Delta\mu_2}{\Gamma}
	p_2\frac{\partial\phi}{\partial X_2}
	\right).
\end{align}
Finally, taking the steady-state average of Eq.~\eqref{eq:S_as_odd_final}, we obtain Eq.~\eqref{eq:Sas} of the main text.

%\nocite{*}

\twocolumngrid

\bibliographystyle{apsrev4-1}
\bibliography{cite_AHE}% Produces the bibliography via BibTeX.

\begin{thebibliography}{61}
\bibitem [{foo()}]{footnote}%
  \BibitemOpen
  \href@noop {} {}\bibinfo {note} {The inequality also implies $\langle\dot{Q}_1\rangle \ge \eta_\mathrm{C}^{-1}(\langle\dot{W}_\mathrm{out}\rangle - \langle\dot{W}_\mathrm{chem}\rangle)$, where the rhs can be negative if $\langle\dot{W}_\mathrm{chem}\rangle$ is sufficiently large. Thus, the active heat engine can extract work even while generating heat ($\langle\dot{Q}_1\rangle < 0$) in the hot reservoir.}\BibitemShut {Stop}%
\end{thebibliography}

\end{document}